\documentclass[journal,twoside,web]{ieeecolor}
\usepackage{tmi}
\usepackage{cite}
\usepackage{amsmath,amssymb,amsfonts}
\usepackage{algorithmic}
\usepackage{graphicx}
\usepackage{textcomp}
\usepackage{amsmath}
\usepackage[table,xcdraw]{xcolor}
\usepackage{booktabs}
\usepackage{multirow}
\usepackage{setspace}
\usepackage{subfigure}
\graphicspath{{figures/},{pics}}
\def\BibTeX{{\rm B\kern-.05em{\sc i\kern-.025em b}\kern-.08em
T\kern-.1667em\lower.7ex\hbox{E}\kern-.125emX}}
\begin{document}
\title{FISTA-Net: Learning A Fast Iterative Shrinkage Thresholding Network for Inverse Problems \\ in Imaging}
\author{Jinxi~Xiang,~\IEEEmembership{Graduate Student Member,~IEEE,}
Yonggui~Dong,
and~Yunjie~Yang,~\IEEEmembership{Member,~IEEE}
\thanks{This work was support in part by National Natural Science Foundation of China under grant 61671270, 62071269. The work of Jinxi Xiang was supported in part by the China Scholarship Council under Grant 201906210259. (Correspondence: Yunjie Yang and Yonggui Dong)}
\thanks{Jinxi Xiang and Yonggui Dong are  with  the Department of Precision Instrument, Tsinghua University, Beijing 100084, China. Jinxi Xiang was a visiting PhD student at the Agile Tomography Group, The University of Edinburgh. (xiangjx16@mails.tsinghua.edu.cn, dongyg@mail.tsinghua.edu.cn).}
\thanks{Yunjie Yang is with the Agile Tomography Group, Institute for Digital Communications, School	of Engineering, The University of	Edinburgh, Edinburgh EH9 3FG, U.K. (y.yang@ed.ac.uk).}
}
\maketitle

\begin{abstract}
Inverse problems are essential to imaging applications.
In this paper, we propose a model-based deep learning network, named FISTA-Net, by combining the merits of interpretability and generality of the model-based Fast Iterative Shrinkage/Thresholding Algorithm (FISTA) and strong regularization and tuning-free advantages of the data-driven neural network. By unfolding the FISTA into a deep network, the architecture of FISTA-Net consists of multiple gradient descent, proximal mapping, and {momentum modules} in cascade.
{Different from FISTA, the gradient matrix in FISTA-Net can  be updated during iteration and a proximal operator network is developed for nonlinear thresholding which can be learned through end-to-end training. Key parameters of FISTA-Net including the gradient step size, thresholding value and momentum scalar are tuning-free and learned from training data rather than hand-crafted.} We further impose positive and monotonous constraints on these parameters to ensure they converge properly.
The experimental results, evaluated both visually and quantitatively, show that the FISTA-Net can optimize parameters for different imaging tasks, i.e. Electromagnetic Tomography (EMT) and X-ray Computational Tomography (X-ray CT). It outperforms the state-of-the-art model-based and deep learning methods and 
{exhibits good generalization ability over  other competitive learning-based approaches under different noise levels.}
\end{abstract}

\begin{IEEEkeywords}
Deep learning, EMT, FISTA, inverse problem, image reconstruction,  model-based method, sparse-view CT
\end{IEEEkeywords}

\section{Introduction}
\label{sec:introduction}

\IEEEPARstart{I}{nverse} problems for imaging applications are at the core of many challenging applications throughout physical and biomedical sciences, including optical and radar systems, Magnetic Resonance Imaging (MRI), X-ray Computed Tomography (CT), Positron Emission Tomography (PET), Ultrasound Tomography (UT), and Electrical Tomography (ET). The goal of an inverse problem in imaging is to estimate an unknown image $\mathbf{x}$ from given measurements $\mathbf{b}$ which relates to $\mathbf{x}$ via a forward operator $\mathbf{A}_{\mathbf{y}}$. The forward model of an imaging problem has the form \cite{Chung2010AnEI}:
\begin{equation}
\mathbf{b} = \mathbf{A}_{\mathbf{y}} (\mathbf{x})+\boldsymbol{\varepsilon}
\label{eq:basic}
\end{equation}
where $\mathbf{A}_{\mathbf{y}}$ denotes the forward operator; $\boldsymbol{\varepsilon}$ is the noise in the measured data.  If $\mathbf{y}$ is independent with $\mathbf{x}$ and  the operator $\mathbf{A}_{\mathbf{y}}$ is linear, it describes the linear inverse problem of imaging systems, e.g. MRI, CT. On the other hand, $\mathbf{A}_{\mathbf{y}}$ represents a non-linear operator, modelling 'soft-field' imaging systems such as ET and Diffuse Optical Tomography (DOT) \cite{Lee2011CompressiveDO}. 

The inversion problem of (\ref{eq:basic}) can be generally formulated as an optimization problem that minimizes the cost function:
\begin{equation}
\underset{\mathbf{x}}{\operatorname{argmin}}\ \mathcal{D}(\mathbf{x}, \mathbf{b})+\lambda \mathcal{R}(\mathbf{x})
\label{eq:reg_basic}
\end{equation}
where $\mathcal{D}$ denotes data fidelity to ensure consistency between the reconstructed image $\mathbf{x}$ and measurements $\mathbf{b}$; $\mathcal{R}$ is a regularizer that imposes  prior knowledge, e.g. smoothness, sparsity, low-rank, non-local self-similarity; $\lambda > 0$ is the regularization coefficient.
Frequently, regularizer $\mathcal{R}$ is non-smooth, which cannot be solved straightforward. First-order iterative  proximal gradient methods, such as the Iterative Shrinkage/Thresholding Algorithm (ISTA) \cite{figueiredo2003algorithm, bioucas2007new, beck2009fast},  Alternating Direction Method of Multipliers (ADMM) \cite{Boyd2011DistributedOA} and Primal-Dual Hybrid Gradient (PDHG) algorithm \cite{Chambolle2010AFP} are prevailing approaches to solve such problems with non-smooth regularizers with high computational efficiency. 
{While ISTA and ADMM have been extensively investigated in solving linear inverse problems, there also exists some variants for nonlinear problems \cite{Kamilov2017APP, Wang2017NonconvexGO, Wei2020TuningfreePP}. Most remarkably, researchers have also proposed computationally tractable approaches to evaluate the gradient of nonlinear forward operators, which can be readily plugged into the FISTA framework \cite{Kamilov2016ARB, Kamilov2017APP}.}


In addition to these model-based approaches, machine learning, particularly deep learning, has recently become a new frontier of inverse problems in imaging \cite{Wang2018ImageRI}. Deep learning is commonly used for non-linear function approximation under weak assumptions. Solving the inverse problem of (\ref{eq:basic}) with deep learning can be phrased as seeking a (non-linear) mapping ${f}_{\Theta}^{\dagger}$ that satisfies the pseudo-inverse property \cite{adler2017solving}:
\begin{equation}
\mathbf{x}_{\text {rec}} = {f}_{\Theta}^{\dagger}(\mathbf{b}) 
\label{eq:DLInverse}
\end{equation}

\begin{figure}[!t]
\centerline{\includegraphics[width=3.3in]{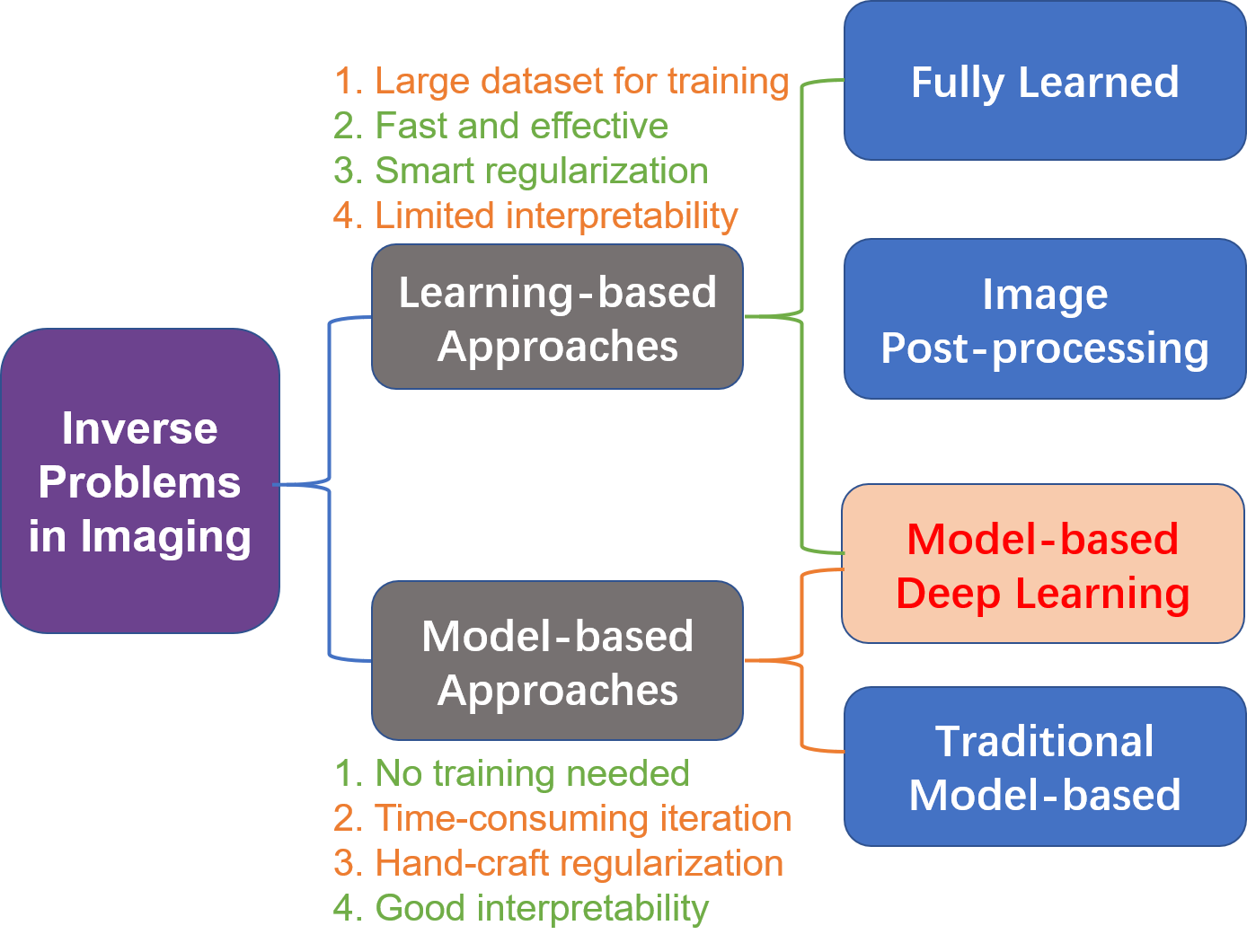}}
\caption{Common approaches for solving inverse problems in imaging.}
\label{fig:introduction}
\end{figure}

There are three common approaches for solving inverse problems with deep learning (see Fig. \ref{fig:introduction}), i.e. 

\subsubsection{Fully learned approach} This approach utilizes conventional neural networks as  'black-box' by feeding sufficiently large amount of training samples to learn the linear/non-linear mapping from the measured data $\mathbf{b}$ to the target image $\mathbf{x}$ without explicitly modeling the domain knowledge, i.e. the forward operator $\mathbf{A}$ \cite{argyrou2012tomographic, tan2018image, Zheng2018AnAI, Xiao2018DeepLI}. In this approach, the deep learning model must learn the underlying physics of the problem, which is difficult or even infeasible \cite{McCann2017ConvolutionalNN}. Thus far, the success of such 
fully data-driven approaches is confined to tasks where the forward operators are insignificant. 
It is challenging for a neural network to invert the process of Eq. (\ref{eq:basic}) from a low-dimension data $\mathbf{b}$ to a high dimensional image $\mathbf{x}$\cite{Yang2020ADMMCSNetAD}. Moreover, their generalization ability has yet to be thoroughly demonstrated in practice\cite{Adler2018LearnedPR}. 

\subsubsection{Image post-processing} This method applies deep learning based image post-processing to suppress noise and remove artifacts. An initial 'coarse' image is first generated by certain analytical inversion, e.g. Filtered Back Projection (FBP)\cite{Jin2017DeepCN}, transforming from measurement domain to image domain by leveraging the forward operator $\mathbf{A}$. A trained deep neural network ${f}_{\mathrm{w}}$ is cascaded afterwards to post process the initial image and obtain the final 'finer' image:
\begin{equation}
\mathbf{x}_{\mathrm{rec}}={f}_{\mathrm{w}}\left(\mathbf{A}^{T} \mathbf{b}\right)
\label{eq:postprocess}
\end{equation}

Notable success of this approach has been achieved especially in low-dose CT imaging \cite{Jin2017DeepCN, han2016deep, Chen2017LowDoseCW, shan2019competitive}. Whilst the images can be improved in terms of certain metrics, e.g. Peak Signal-to-Noise Ratio (PSNR), some researchers pointed out that the 'cosmetic improvements' might result in over-smoothing and loss of critical structural details in certain cases \cite{Yang2018LowDoseCI}. The performance of ${f}_{\mathrm{w}}$ is limited by the information content of initial images. Without a feedback mechanism that imposes data consistency, large networks requiring a considerable amount of training data are often needed in practical applications \cite{Aggarwal2019MoDLMD}.

\subsubsection{Model-based deep learning} The forward operator $\mathbf{A}$ encapsulates the relevance of the measured data $\mathbf{b}$ and the target image $\mathbf{x}$. With the forward operator depicting the underlying physics, conventional model-based algorithms are underpinned by two pillars, i.e. generality and stability. On the other hand, despite the effectiveness of deep learning in capturing critical features of data, its stability is usually unsatisfactory \cite{Antun2020OnIO}. The model-based deep learning approach aims to combine the advantages of these two paradigms.
{Model-based deep learning offers a variational architecture for solving the inverse problem with a physical model describing data generation, a statistical model revealing the noise pattern in data, and a prior model promoting desired features. The forward operator and the statistical model ensures the consistency of the image against the measured data, which could considerably improve the generalization ability\cite{hammernik2018learning, Maier2019LearningWK}.}

This idea was first proposed in the Learned Iterative Shrinkage-Thresholding Algorithm (LISTA) \cite{Gregor2010LearningFA}. It aims to train a non-linear predictor through unrolling the iterative algorithm into feed-forward layers, which are optimized in a data-driven manner. Following this seminal work, many similar algorithms have been developed recently to solve the inverse problems in imaging. For instance,
Yang \textit{et al.} proposed a deep architecture, dubbed ADMM-CSNet, by unrolling ADMM solver into iterative steps, and all parameters are discriminatively learned by end-to-end training for compressed sensing applications\cite{Yang2020ADMMCSNetAD}.
Zhang \textit{et al.} cast ISTA into deep network and developed an effective strategy to solve the proximal mapping using nonlinear transforms \cite{Zhang2018ISTANetIO}. 
Adler \textit{et al.} proposed an iterative Convolutional Neural Network (CNN) to learn the gradient descent \cite{adler2017solving, Adler2018LearnedPR}. Hammernik \textit{et al.} embedded a variational model in an unrolled gradient descent scheme and all parameters of the prior model
are learned during the training procedure \cite{hammernik2018learning}. Gong \textit{et al.} formulated the inverse problem using ADMM with CNN represented prior \cite{Gong2017IterativePI}. Aggarwal \textit{et al.} proposed MoDL, a variational framework involving a data-consistency term and a learned CNN to capture the image redundancy \cite{Aggarwal2019MoDLMD}. All referred work acknowledged the significance of embedding the data consistency term in the deep learning network, which improved significantly its generalization ablity and convergence performance.

This paper aims to ground the model-based deep learning approach on the Fast Iterative Shrinkage/Thresholding Algorithm (FISTA) \cite{beck2009fast} by unfolding the iterative step into cascaded blocks and replacing the nonlinear soft-thresholding function with a deep network, named as FISTA-Net. 
Main advantages of FISTA-Net include:

\begin{enumerate}
\item Compared with the state-of-the-art ADMM-based networks, e.g. ADMM-CSNet\cite{Yang2020ADMMCSNetAD}, FISTA-Net does not involve matrix inversion of the forward operator, which is desirable when dealing with large-scale or ill-conditioned inverse problems\cite{Kamilov2017APP}. 
\item A CNN is designed to solve the proximal mapping associated with the sparsity-constrained regularizer.  The learned CNN is shared throughout the iterations with different thresholding values to deal with varying noise at different iteration. The network is much easier to train with a smaller size compared with existing methods.
\item FISTA requires manual parameter tweaking  (e.g. gradient step size, regularization parameters) for different scenarios. In contrast, FISTA-Net autonomously determines these parameters from the data manifold. In addition, model-based constraints are further imposed on these parameters to ensure they converge properly.
\end{enumerate}

{The major differences between the state-of-the-art methods (e.g. ISTA-Net\cite{Zhang2018ISTANetIO}) and FISTA-Net are: 
(1) In FISTA-Net, the gradient matrix is learned by substituting the fixed classic gradient  throughout iterations. Unfolding iterations lead to better reconstruction if the weights are allowed to vary at each layer\cite{Gregor2010LearningFA}.
(2)	The core parameters of FISTA-Net, e.g. the step size, thresholding values, are regularized to converge properly. This is derived from the observation that ISTA-Net could come with non-positive step sizes and thresholding values, which contradicts their definitions. In FISTA-Net, the variables are regularized to be positive and change gradually with noise level throughout iterations.
(3) A momentum module is added in FISTA-Net to accelerate convergence. Momentum is widely recognized as a means of dampening oscillations and speeding up iterations, which will lead to faster convergence\cite{beck2009fast, Moreau2017UnderstandingTS}.}

In the following Sections, we first describe the FISTA-Net starting from the original FISTA framework and then cast it into the deep learning network. Then, we test and evaluate FISTA-Net on two different imaging modalities, i.e. CT and Electromagnetic Tomography (EMT).

\section{Iterative Shrinkage/Thresholding Frameworks}

\subsection{ISTA}
Iterative Shrinkage/Thresholding Algorithm (ISTA) is a prevailing framework to solve (\ref{eq:reg_basic}) with non-smooth convex regularizers. Each iteration of ISTA involves gradient descent update followed by a shrinkage/soft-threshold step (e.g. proximal operator)\cite{Daubechies2003AnIT, Elad2006WhySS}:
\begin{equation}
\mathbf{x}^{(k+1)}=\mathcal{T}_{\alpha }\left(\mathbf{x}^{(k)}-\mu \mathbf{A}^{T}\left(\mathbf{A} \mathbf{x}^{(k)}-\mathbf{b}\right)\right)
\label{eq:ista}
\end{equation}
where $\mu=1/L$ is an appropriate step size, and $L$  must be an upper bound on the largest eigenvalue of $\mathbf{A}^T\mathbf{A}$, e.g. $L>\lambda_{\text{max}}(\mathbf{A}^T\mathbf{A})$ \cite{Gregor2010LearningFA}. { $\mathcal{T}_\alpha$ is the shrinkage/thresholding operator associated with the regularizer $\lambda\mathcal{R}(\mathbf{x})$ which is defined in a component-wise way. }
For certain $\mathcal{R}(\mathbf{x})$, $\mathcal{T}_{\alpha}$ has closed forms. For instance, $\mathcal{T}_{\alpha}(\mathbf{x}) = \text{soft}(\mathbf{x}, \alpha)$ when $\mathcal{R}(\mathbf{x}) = \|\mathbf{x}\|_{1}$, where $\text{soft}(\mathbf{x}, \alpha) = \text{sign}({x}_i)\ \text{max}\{|x_i|-\alpha, 0\}$. The shrinkage/threshold function of $\mathcal{R}(\mathbf{x}) = \|\mathbf{x}\|_{0}$ is  $\mathcal{T}_{\alpha}(\mathbf{x}) = \text{hard}(\mathbf{x}, \alpha)$, where $\text{hard}(\mathbf{x_i}, \alpha) = x_i 1_{|x_i| \geq \alpha}$. A comprehensive coverage of proximal maps can be found in\cite{Combettes2005SignalRB}. 

\subsection{FISTA}

ISTA is generally recognized as a time-consuming  method\cite{beck2009fast, bioucas2007new}.  Two faster versions of ISTA are the two-step IST (TwIST) algorithm\cite{bioucas2007new, Afonso2010FastIR},
and fast IST algorithm (FISTA)\cite{beck2009fast}. 
FISTA solves (\ref{eq:reg_basic})  by iterating the following update steps:
\begin{subequations}
\begin{align}
	\label{subeq:fista_a}
	\mathbf{x}^{(k)}&=\mathcal{T}_{\alpha }\left(\mathbf{y}^{(k)}-\mu \mathbf{A}^{T}\left(\mathbf{A} \mathbf{y}^{(k)}-\mathbf{b}\right)\right) \\
	\label{subeq:fista_b}
	t^{(k+1)} &=\frac{1+\sqrt{1+4 (t^{(k)})^{2}}}{2} \\
	\label{subeq:fista_c}
	\mathbf{y}^{(k+1)} &=\mathbf{x}^{(k)}+\left(\frac{t^{(k)}-1}{t^{(k+1)}}\right)\left(\mathbf{x}^{(k)}-\mathbf{x}^{(k-1)}\right)
\end{align}
\label{eq:fista}
\end{subequations}

The main improvement of FISTA is that the iterative shrinkage operator $\mathcal{T}_{\alpha }$ is not applied on the previous estimation $\mathbf{x}^{(k)}$, but rather at $\mathbf{y}^{(k)}$ which adopts a well-designed linear combination of the previous two estimates $\mathbf{x}^{(k)},\ \mathbf{x}^{(k-1)}$. FISTA does not require more than one gradient evaluation at each iteration but just an additional estimation that is smartly chosen.

\section{FISTA-Net}

Sparse representation, lying on the fact that most natural images could be well described by fewer basis coefficients in certain transform domains, is an important tool in imaging\cite{GuerquinKern2011AFW}. Herein we define a transform-domain sparsity regularizer $\mathcal{R}(\mathbf{x}) = \|\mathcal{F}(\mathbf{x})\|_{1}$, and rewrite (\ref{eq:reg_basic}) as:
\begin{equation}
\hat{\mathbf{x}}_{\mathcal{F}_{l1}}=\underset{\mathbf{x}}{\operatorname{argmin}}\left\{\|\mathbf{A} \mathbf{x}-\mathbf{b}\|^{2}_2+\lambda\|\mathcal{F}(\mathbf{x})\|_{1}\right\}
\label{eq:x_F1}
\end{equation}
where the operator $\mathcal{F}(\cdot)$ could be Fourier Transform (FT),  Discrete Cosine Transform (DCT) or Discrete Wavelet Transform (DWT). Among them, DWT is the most common sparse transform which has been applied in various fields\cite{vonesch2008fast, zhang2015exponential}.

\subsection{Network Mapping of FISTA}

\begin{figure*}[!t]
\centerline{\includegraphics[width=7in]{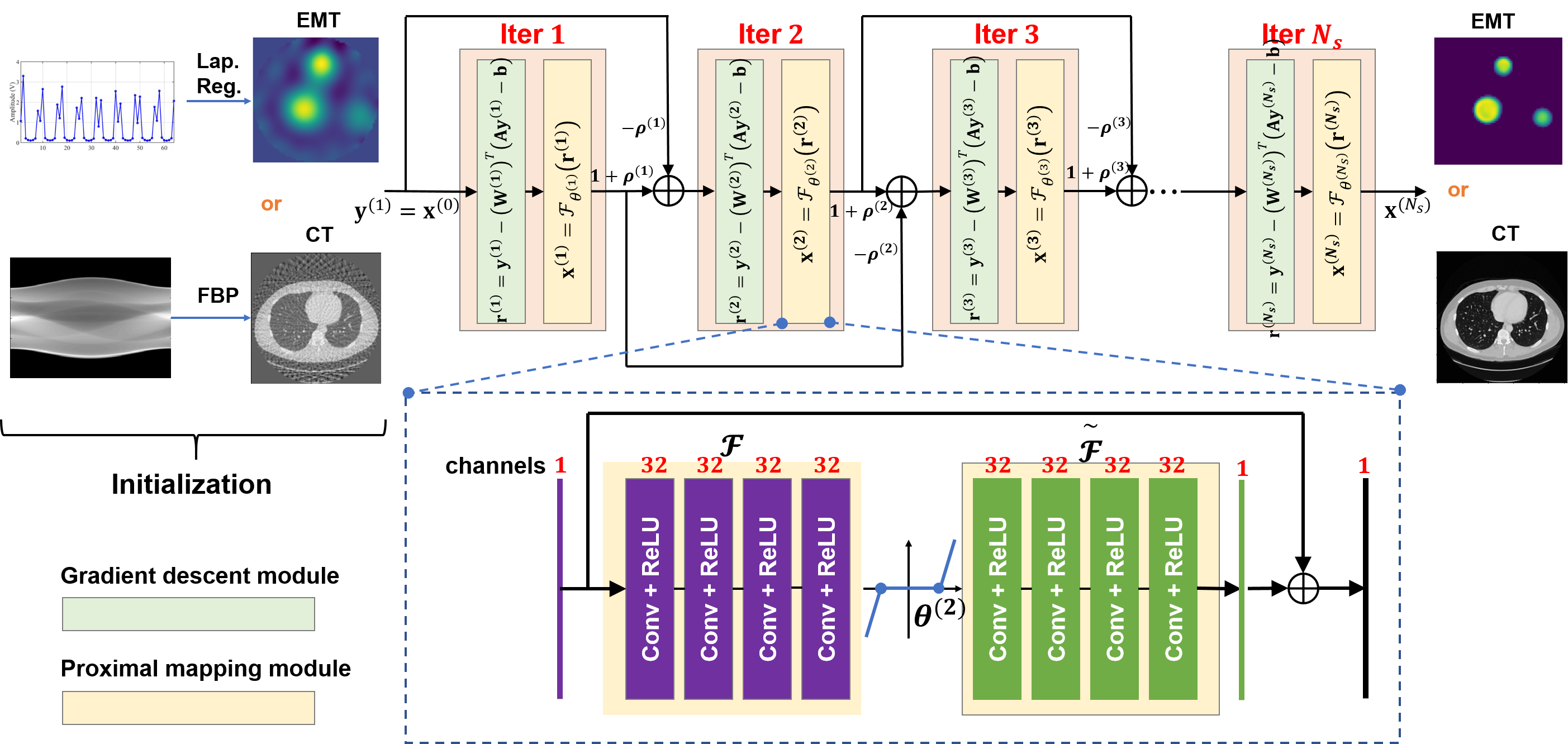}}
\caption{The overall architecture of the proposed FISTA-Net with $N_s$ iterations. In specific, FISTA-Net consists of three main modules, i.e. gradient descent, proximal mapping and two-step update.}
\label{fig:FISTANet}
\end{figure*}

The proposed FISTA-Net to solve (\ref{eq:x_F1}) is formulated as:
\begin{subequations}
\begin{align}
	\label{subeq:gradient}
	\mathbf{r}^{(k)}&=\mathbf{y}^{(k)} - (\mathbf{W}^{(k)})^T \left(\mathbf{A} \mathbf{y}^{(k)}-\mathbf{b}\right) \\ 
	\label{subeq:threshold}
	\mathbf{x}^{(k)} &= \mathcal{T}_{\theta^{(k)}}(\mathbf{r}^{(k)})\\
	\label{subeq:twostep}
	\mathbf{y}^{(k+1)} &=\mathbf{x}^{(k)} + {\rho}^{(k)}  \left(\mathbf{x}^{(k)}-\mathbf{x}^{(k-1)}\right)
\end{align}
\label{eq:FISTA_Net}
\end{subequations}
{where $\mathbf{r}^{(k)}$, $\mathbf{y}^{(k)}$, and $\mathbf{x}^{(k)}$ are intermediate variables of the final result; $\mathbf{W}^{(k)}$ is the  gradient operator; $\mathcal{T}_{\theta^{(k)}}$ denotes the  nonlinear proximal operator; ${\rho}^{(k)}$ represents the  scalar for momentum update.}

{FISTA-Net alternates between the minimization of the smooth differentiable part using the gradient information in (\ref{subeq:gradient}), and the minimization of the non-differentiable part using a proximal operator represented by a learned network in (\ref{subeq:threshold}).} The update in (\ref{subeq:twostep}) is a two-step linear combination of the previous two iterations.  Fig. \ref{fig:FISTANet} illustrates the overall architecture of FISTA-Net and more details are provided hereafter.

\textbf{Gradient descent module $\mathbf{r}^{(k)}$.}
This layer updates the reconstructed image based on the gradient descent operation of Eq. (\ref{subeq:gradient}), which is the closed-form numerical solution of the data consistency subproblem, given $\mathbf{y}^{(k)}$ as the output of the previous layer. Explicitly, it aims to find a more accurate estimation that minimizes $\|\mathbf{A} \mathbf{x}-\mathbf{b}\|^{2}_2$. As the data consistency term corresponds to the physical model, it imposes a physics constraint to stabilize the solution.
{In\cite{Maier2019LearningWK}, researchers provide a} theoretical discussion on learning with known forward operators, which can reduce the maximum error bounds.

{For Gaussian noises problems ($\boldsymbol{\varepsilon}$ is a Gaussian vector in (\ref{eq:basic})) with $\mathcal{R}(\mathbf{x}) = \|\mathbf{x}\|_{1}$, the formulation of (\ref{subeq:gradient}) reduces to (\ref{subeq:fista_a}) by letting  $\mathbf{W}^{(k)} = \mu \mathbf{A}$\cite{Wu2020SparseCW}. Inspired by ISTA in (\ref{eq:ista}), the Learned ISTA (LISTA)\cite{Gregor2010LearningFA} proposed to learn the weights in the matrices in ISTA rather than fixing them:}
\begin{equation}
\mathbf{x}^{(k+1)}=\mathcal{T}_{\alpha }\left(\mathbf{W}_{1}^{(k)} \mathbf{b}+\mathbf{W}_{2}^{(k)} \mathbf{x}^{(k)}\right)
\label{eq:LISTA}
\end{equation}

{And the necessary condition for LISTA convergence is\cite{Chen2018TheoreticalLC}:
\begin{equation}
	\mathbf{W}_{2}^{(k)}-\left(\mathbf{I}-\mathbf{W}_{1}^{(k)} \mathbf{A}\right) \rightarrow \mathbf{0}, \quad \text { as } k \rightarrow \infty
	\label{eq:partial_weights}
\end{equation}}

{Eq. (\ref{subeq:gradient}) satisfies the necessary condition of convergence by letting $\mathbf{W}_{1}^{(k)}=\left(\mathbf{W}^{(k)}\right)^{T}, \mathbf{W}_{2}^{(k)}=\mathbf{I}-\mathbf{W}_{1}^{(k)} \mathbf{A}$. The weight $\mathbf{W}^{(k)}$ of (\ref{subeq:gradient}) is a linear operator with the same dimensionality with $\mathbf{A}$. Liu \textit{et al.} \cite{Liu2019ALISTAAW} showed that $\mathbf{W}^{(k)}$ can be decomposed as the product of a scalar $\mu^{(k)}$ and a matrix $\tilde{\mathbf{W}}$ independent of layer index $k$:
\begin{equation}
	\mathbf{W}^{(k)} = \mu^{(k)} \tilde{\mathbf{W}}
\end{equation}
where $\tilde{\mathbf{W}}$ has small coherence with $\mathbf{A}$. The matrix $\tilde{\mathbf{W}}$ is pre-computed by solving\cite{Liu2019ALISTAAW}:
\begin{equation}
	\begin{array}{l}
		\tilde{\mathbf{W}} \in \underset{\mathbf{W} \in \mathbb{R}^{N \times M}} {\arg \min }\left\|\mathbf{W}^{T} \mathbf{A}\right\|_{F}^{2} \\
		\text { s.t. }\left(\mathbf{W}_{:, m}\right)^{T} \mathbf{A}_{:, m}=1, \forall m=1,2, \cdots, M
	\end{array}
	\label{eq:analytical_W}
\end{equation}}

{The objective of (\ref{eq:analytical_W}) is to minimize the Frobenius norm of $\mathbf{W}^T\mathbf{A}$ over linear constraints. This is a standard convex quadratic program that can be solved efficiently\cite{Boyd2006ConvexO}. With $\tilde{\mathbf{W}}$ being solved before-hand, the step size $\mu^{(k)}$ is learned through end-to-end training.}

\textbf{Proximal mapping module $\mathbf{x}^{(k)}$.}
The proximal operator aims to remove noise and artifacts in the intermediate result $\mathbf{r}^{(k)}$ through thresholding in a certain transform domain. In practice, existing sparse transformations need fine-tuning to capture complex image details. FISTA-Net aims to learn a more flexible representation $\mathcal{T}(\cdot)$ and threshold $\theta^{(k)}$ from training data. 

In FISTA-Net, $\mathcal{T}(\cdot)$ is designed as a combination of four linear convolutional operators (without bias terms) separated by a Rectified Linear Unit (ReLU) (see the dashed box in Fig. \ref{fig:FISTANet}). The first convolutional operator corresponds to $N_f$ filters (each of size $3\times3$) and the other three convolutional layers correspond to $N_f$ filters (each of size $3 \times 3 \times N_f$). We set $N_f = 32$ by default. In order to inherit the merits of residual learning, a skip connection is made from input to output. Batch Normalization (BN) is not adopted because some recent studies show that BN layers are more likely to introduce unpleasant artifacts when the network becomes deeper and more complicated \cite{wang2018esrgan, zhang2018residual}. 

Mathematically, the sparse transform is invertible, i.e. $\widetilde{\mathcal{F}} \circ \mathcal{F}=\mathcal{I}$,  where  $\mathcal{I}$ is the identity operator. Inspired by the loss function of sparse autoencoder\cite{goodfellow2016deep}, we define the loss of the transform $\mathcal{F}(\cdot)$ as:
\begin{equation}
\begin{aligned}
	\mathcal{L}_{\text {tsf}} & = \lambda_1 \mathcal{L}_{\text {sym}} + \lambda_2 \mathcal{L}_{\text {spa}}\\
	& =\lambda_1 \sum_{k=1}^{N_{s}}\left\|\widetilde{\mathcal{F}}\left(\mathcal{F}\left(\mathbf{r}^{(k)}\right)\right)-\mathbf{r}^{(k)}\right\|_{2}^{2} + \lambda_2 \sum_{k=1}^{N_{s}} \|\mathcal{F}(\mathbf{r}^{(k)})\|_{1}
\end{aligned}
\label{eq:loss_sym}
\end{equation}
where $\mathcal{L}_{\text {sym}}$ is a symmetry loss of inversion; $\mathcal{L}_{\text {spa}}$ is a sparse constraint.

To deal with the changing noise/artifact level at each iteration, previously reported plug-and-play CNN approaches require various pretrained denoiser with different noise levels\cite{Zhang2017ImageRF, Zhang2017LearningDC}. MoDL shares the same network and the same regularization parameter across iterations regardless of the variance of the noise level\cite{Aggarwal2019MoDLMD}. In FISTA-Net, differently, the parameters of $\mathcal{T}(\cdot)$ are shared throughout different iterations, whereas $\theta^{(k)}$, the shrinkage thresholding value, is a learnable parameter. As the variance of noise and artifact progressively changes with iterations, we allow $\theta^{(k)}$ to 
change at each cascaded stage. The advantage of such setting is that we could maintain the flexibility to adapt the noise variance at each iteration while avoiding training various networks.

\textbf{Momentum module $\mathbf{y}^{(k+1)}$.} The original ISTA can be accelerated by introducing a momentum term\cite{Nesterov2005SmoothMO}, as it is done in FISTA that is proven to converge in function values as $O\left(1 / k^{2}\right)$ compared to the slower rate of $O\left(1 / k\right)$  of ISTA\cite{beck2009fast}. Its faster convergence rate than ISTA lies in smartly choosing the update weights of two previous results without requiring additional gradient evaluation. In FISTA-Net, we inherit this advantage by relaxing the constant update weights $\{ t^{(k)}\}$ of (\ref{subeq:fista_c}) with a learnable parameter $\rho^{(k)}$ that is autonomously learned from the training dataset.

The total loss function of FISTA-Net is formulated as:
\begin{equation}
\mathcal{L}_{\text {total}} = \mathcal{L}_{\text {mse}} + \lambda_1 \mathcal{L}_{\text {sym}} + \lambda_2 \mathcal{L}_{\text {spa}} 
\label{eq:total_loss}
\end{equation}
where $\left\|\mathbf{x}_{N_s}-\mathbf{x}_{gt}\right\|_{2}^{2}$ is the MSE loss of the estimated output of FISTA-Net with respect to the ground truth. We train the network for different tasks with fixed  hyperparameters, i.e. $\lambda_1=0.01, \ \lambda_2=0.001$ by default. $\lambda_2$ is set to a smaller value because the thresholding function has a similar effect. Empirically,  we found it is adequate to set these weights such that the magnitude of different loss terms is balanced into similar scales.

\subsection{Model-based Parameter Constraints}
Although $\{\mu^{(k)}, \theta^{(k)}, \rho^{(k)} \}_{k=1}^{N_s}$ are learnable and no manual parameter tweaking is required in FISTA-Net, we introduce extra constraints to ensure they  converge properly. 
{This is based on the observations that ISTA-Net could possibly generate non-positive step size and thresholding values during iteration, which contradicts the definition of these variables.}
Therefore, a good rule of thumb is to let $\{\mu^{(k)}, \theta^{(k)}, \rho^{(k)} \}_{k=1}^{N_s}$  be positive. In addition, the gradient step $\mu_k$ should decay smoothly with iterations. Thresholding value $\theta^{(k)}$ should also iteratively decrease because the noise variances are suppressed progressively with iterations. The two-step update weight $\rho^{(k)} $ should increase monotonously corresponding to the two-step update weight in the FISTA. Henceforth, we regularize:
\begin{subequations}
\begin{align}
	\label{subeq:sp_mu}
	\mu^{(k)} &= {sp} (w_1 k + c_1),\ \ w_1<0  \\ 
	\label{subeq:sp_theta}
	\theta^{(k)} &= {sp} (w_2 k + c_2),\ \ w_2<0 \\
	\label{subeq:sp_rho}
	\rho^{(k)} &= \frac{{sp} (w_3 k + c_3) - {sp} (w_3+ c_3)}{ {sp} (w_3 k + c_3)},\ \ w_3>0
\end{align}
\label{eq:softplus}
\end{subequations}
where the softplus function ${sp}(x) = \ln (1+\exp(x)) $; $\rho^{(k)}\in (0,1)$ is consistent with FISTA; the iteration step $k=1,2,\dots,N_s$. One benefit of the softplus function is its simple derivative function.

\subsection{Initialization}

ISTA-based methods can benefit significantly from warm-starting, i.e. initialization near a minimum of the objective function. For ill-conditioned problems, e.g. EMT, we adopt the solution of (\ref{eq:reg_basic}) with Laplacian operator $\mathcal{R}(\mathbf{x}) = \| \mathbf{L}\mathbf{x}\|^2_2$, ($\mathbf{L}$ is the Laplacian matrix)\cite{Yang2017AME, Xiang2019DesignOA} to initialize FISTA-Net:
\begin{equation}
\mathbf{x}^{(0)} = (\mathbf{A}^T\mathbf{A} + \lambda_0\mathbf{L}^T\mathbf{L})^{-1}\mathbf{A}^T\mathbf{b}
\label{eq:x_lap}
\end{equation}
where we set $\mathbf{y}^{(1)}=\mathbf{x}^{(0)}$, and $\lambda_0=0.001$. 

For well-conditioned problems, e.g. CT, no regularization is needed for initialization and we simply apply \textit{iradon transform} $ \mathbf{x}_0 = \mathbf{A}^T\mathbf{b}$ using Python\cite{Ronchetti2020TorchRadonFD} or Matlab function. In experiments, we found that providing FISTA-Net with such an initial guess marginally decreased the training time, although it did not produce a better final result. 

The convolutional network is initialized with Xavier algorithm \cite{Glorot2010UnderstandingTD}. The parameters $\{w_1, w_2, w_3, c_1, c_2, c_3 \}$ are initialized as $\{-0.5, -0.2, 1, -2, -1, 0\}$.

\subsection{Implementation Details}
The unkown parameters of FISTA-Net are $\Theta=\left\{\mu^{(k)}, \rho^{(k)}, \theta^{(k)} \right\}_{k=1}^{N_{\text{s}}},\  \mathcal{F}(\cdot),\widetilde{\mathcal{F}}(\cdot)$, where $N_s$ is the number of iteration steps, given the training data pairs $\left\{\left(\mathbf{b}^{(i)}, \mathbf{x}^{(i)}\right)\right\}_{i=1}^{N_{t}}$, $N_t$ is size of the training dataset. The iteration step of FISTA-Net is fixed while conducting the end-to-end training. Since the networks are shared throughout and $\{w_1, w_2, w_3, c_1, c_2, c_3 \}$ are decoupled with iteration, we may choose a different number of iterations for reconstruction.
In our experiments, all the networks were optimized using Adam algorithm with a mini-batch size of 64\cite{Kingma2015AdamAM}. Two learning rates $lr_1$ and $lr_2$ are  estimated from the first and second moments of the gradients and applied to the convolutional network and $\{w_1, w_2, w_3, c_1, c_2, c_3 \}$, respectively. The networks were implemented in Python with the Pytorch library and the training of FISTA-Net was performed on a workstation with 4 RTX 2080Ti GPUs. 

\section{Experiments and Results}
\label{sec:expriments}

In this section, we verify the proposed FISTA-Net with two representative imaging modalities, i.e. the non-linear Electromagnetic Tomography (EMT) and the linear sparse-view CT. Three prevailing metrics, i.e. the Root Mean Square Error (RMSE), Peak Signal to Noise Ratio (PSNR) and Structural Similarity Index Measure (SSIM) \cite{Wang2004ImageQA}, are utilized to quantify and evaluate image quality.

\subsection{Nonlinear Case: EMT}

\subsubsection{EMT Dataset}

\begin{figure}[tbp]
\centerline{\includegraphics[width=3.3in]{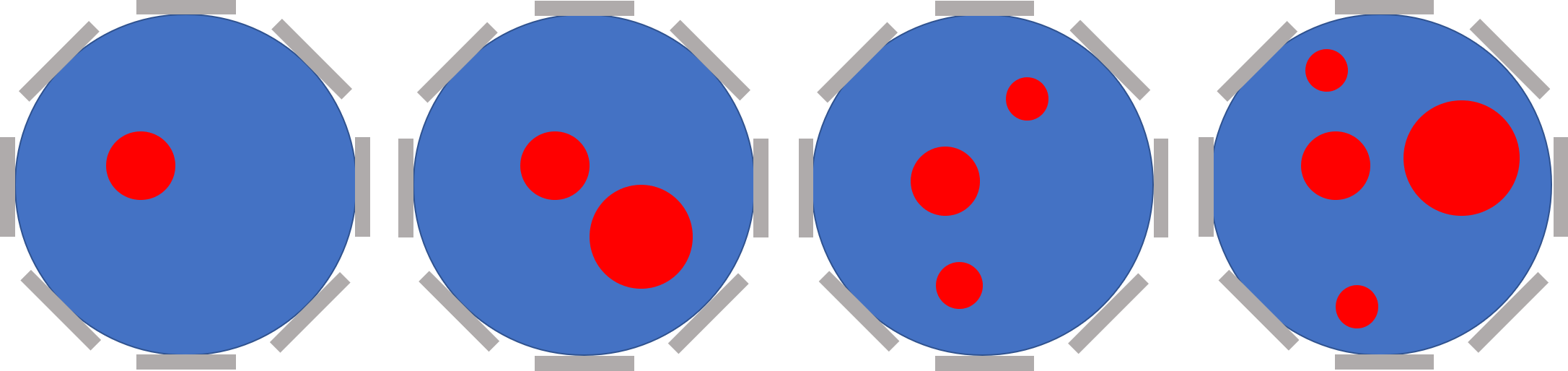}}
\caption{Examples of EMT dataset with different conductivity distribution phantoms. The blue background indicates the sensing region of EMT, and the grey objects represent 8 coils placed in its periphery.}
\label{fig:emt_circles}
\end{figure}

EMT is a non-intrusive and non-radiative tomography modality that can image the cross-sectional conductivity distribution inside the object without contact \cite{Xiang2019DesignOA}. EMT is a nonlinear and ill-conditioned problem caused by the dispersed nature of the electromagnetic field. {It is recognised as a promising imaging technique for biomedical applications, such as detecting the acute stroke \cite{Xiang2020MultifrequencyET}. Stroke lesions or other biological tissues, e.g. tumor cells in \textit{in vitro} disease models\cite{Yang2019ScaffoldBased3C}, are approximately circular. To mimic such imaging objects, we establish an EMT dataset containing multiple circular phantoms with randomly assigned location, conductivity and size. We apply COMSOL Multiphysics to solve the forward problem of EMT \cite{tan2018image, Hu2019ImageRF}.} 
The EMT dataset consists of 24,000 training samples (20\% for validation) and 12,000 testing samples. Each sample contains a 2-D conductivity distribution image  with a dimension of $64\times64$ and a corresponding 1-D measurement vector with 64 elements. In training and validation data, samples contain 1, 2 and 4 round objects with random locations and radius within the sensing region (see Fig. \ref{fig:emt_circles}). The testing data comprise two sets, i.e. set 1 with 1,2,4 round objects and set 2 with 3 round objects to verify the generalization ability. The conductivity of all objects is assigned randomly in the range of $(0.05, 0.5)S/m$. To make data augmentation, white noises are manually added to both training and validation dataset with the SNR of 40dB.

\subsubsection{Comparison Study}

We compare FISTA-Net with four recently reported methods, i.e. Laplacian regularization\cite{Yang2017AME, Xiang2019DesignOA},  FISTA-TV\cite{Beck2009FastGA}, FBPConvNet\cite{Jin2017DeepCN} and {ISTA-Net\cite{Zhang2018ISTANetIO}}. The last two methods and FISTA-Net are trained using pairs of measurement data and ground truth images as input and output, respectively. Laplacian regularization and FISTA-TV are model-based methods.  FBPConvNet is a network-based method. Laplacian regularization is a baseline method that generates the initial guess $\mathbf{x}_0$ for FBPConvNet, ISTA-Net and FISTA-Net. 
{FISTA-TV, derived from\cite{Beck2009FastGA} with necessary modifications, denotes the traditional model-based FISTA method using Total Variation (TV) as the regularizer, i.e. $\mathcal{R}(\mathbf{x}) = \| \mathbf{x}\|_{\text{TV}}$.} 
FBPConvNet is an image post-processing method that suppresses image noise and artifacts with the U-Net denoiser. {ISTA-Net is one of the state-of-the-art model-based deep learning methods originally designed for compressive sensing. }

Table \ref{tab:fig_emt} shows the EMT imaging results of different methods. We can observe that FBPConvNet and FISTA-Net produce much better visual results than traditional model-based methods, with sharper edges and better shapes. It is not surprising that FBPConvNet, which trained a large U-Net with many parameters (note that FISTA-Net is considerably small with only 74,599 learnable parameters, compared to the 482,449 parameters of FBPConvNet) to remove noise, can yield better visual results than ISTA-Net. However, in some cases (row 2, 3), FBPConvNet fails to recover all objects. Further comparing the average quantitative metrics, i.e. PSNR, SSIM and RMSE, on all test data (see Table \ref{tab:metric_emt}), we have the following observations. First, FISTA-Net outperforms other methods on both test sets. {Consistent with visual images, the network-based FBPConvNet performs well in terms of SSIM and is comparable to or even slightly better than FISTA-Net in shape preservation, but it has moderate PSNR and RMSE results. This might be attributed to the strong learning ability of the network to capture image features, whilst RMSE and PSNR cannot be guaranteed without the data fidelity term.} 
Second, all methods work better on set 1 than set 2 because test set 2 with 3 round objects does not appear in the training set.

\begin{table*}[!t]
\caption{Comparison of the proposed FISTA-Net with state of the art parallel imaging approaches on EMT dataset. (First two rows: testing set 1; last two rows: testing set 2)}
\begin{spacing}{1}
	\centering
	\begin{tabular}{ccccccc}
		Ground Truth & Lap. Reg. ($\mathbf{x_0}$) & FISTA-TV\cite{Beck2009FastGA} & FBPConvNet\cite{Jin2017DeepCN} & {ISTA-Net\cite{Zhang2018ISTANetIO}} & \textbf{FISTA-Net} & \\
		\includegraphics[width=2.5cm]{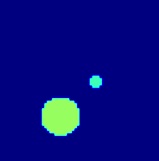} & 
		\includegraphics[width=2.5cm]{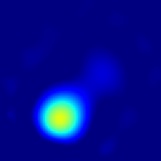} & 
		\includegraphics[width=2.5cm]{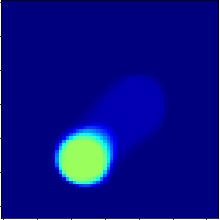} &
		\includegraphics[width=2.5cm]{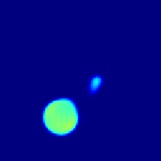} &
		\includegraphics[width=2.5cm]{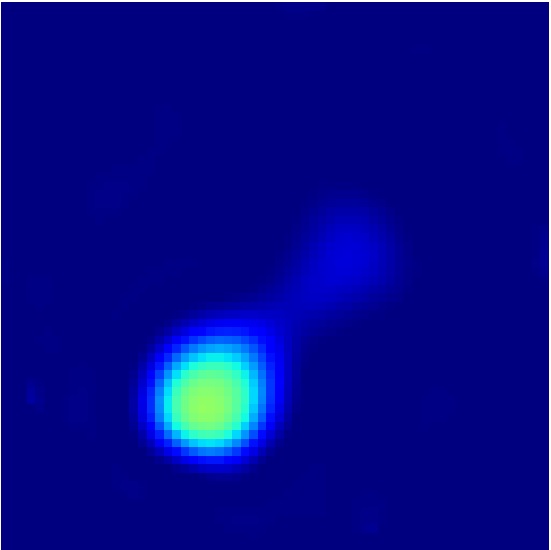} & 
		\includegraphics[width=2.5cm]{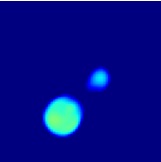} & 
		\includegraphics[height=2.5cm]{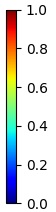} \\
		\includegraphics[width=2.5cm]{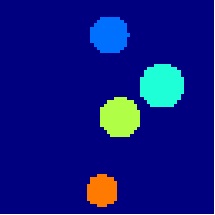} & 
		\includegraphics[width=2.5cm]{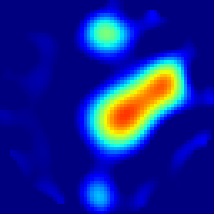} & 
		\includegraphics[width=2.5cm]{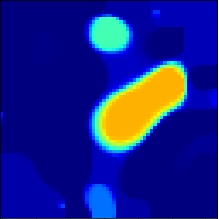} &
		\includegraphics[width=2.5cm]{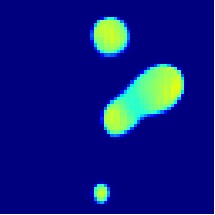} &
		\includegraphics[width=2.5cm]{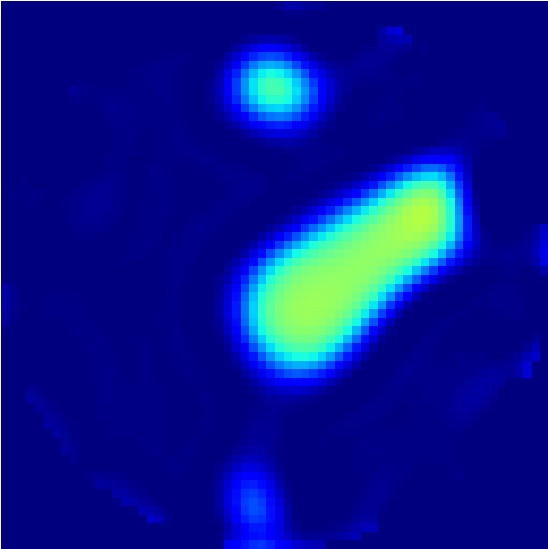} & 
		\includegraphics[width=2.5cm]{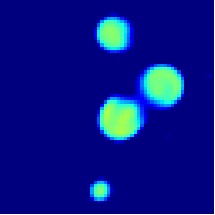} & 
		\includegraphics[height=2.5cm]{cb1.JPG} \\  
		\includegraphics[width=2.5cm]{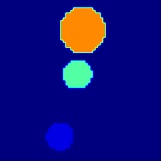} & 
		\includegraphics[width=2.5cm]{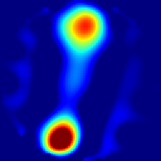} & 
		\includegraphics[width=2.5cm]{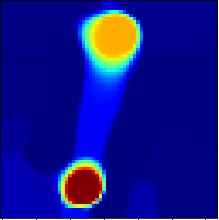} &
		\includegraphics[width=2.5cm]{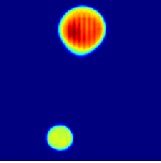} &
		\includegraphics[width=2.5cm]{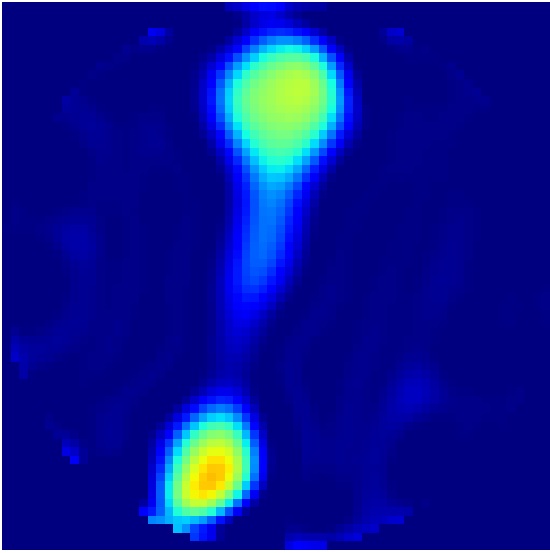} & 
		\includegraphics[width=2.5cm]{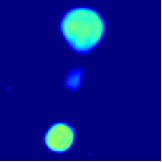} & 
		\includegraphics[height=2.5cm]{cb1.JPG} \\  
		\includegraphics[width=2.5cm]{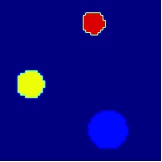} & 
		\includegraphics[width=2.5cm]{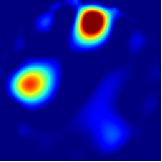} & 
		\includegraphics[width=2.5cm]{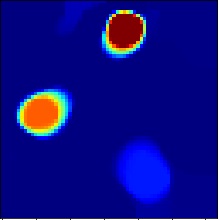} &
		\includegraphics[width=2.5cm]{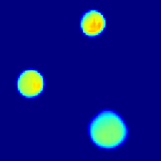} &
		\includegraphics[width=2.5cm]{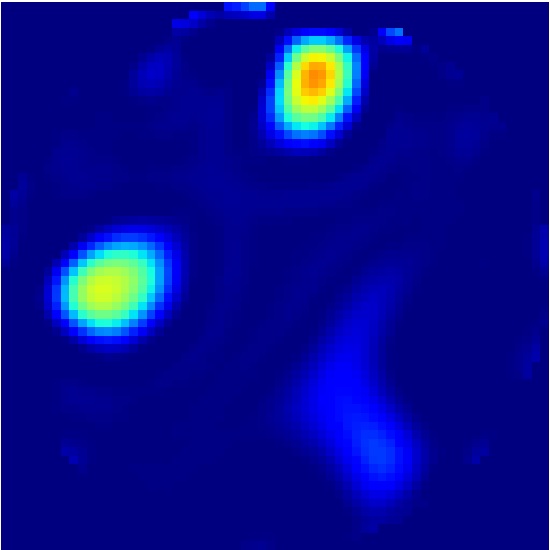} & 
		\includegraphics[width=2.5cm]{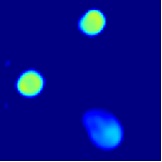} & 
		\includegraphics[height=2.5cm]{cb1.JPG} \\  
	\end{tabular}
\end{spacing}
\label{tab:fig_emt}
\end{table*}

\begin{table*}[!t]
\caption{{Quantitative metrics of different methods on EMT Dataset. The best results of comparative methods are highlighted in \textcolor{blue}{blue}}.}
\centering{
	\begin{tabular}{@{}cm{1.5cm}m{2.5cm}m{2.5cm}m{2.7cm}m{2.2cm}m{2cm}@{}}
		\toprule
		Testing set & Index	 & Lap. Reg. ($\mathbf{x_0}$) & FISTA-TV\cite{Beck2009FastGA} & FBPConvNet\cite{Jin2017DeepCN} & ISTA-Net\cite{Zhang2018ISTANetIO} & \textbf{FISTA-Net}   \\ \midrule
		& \# of Pars. & 1 & 1 & 482449 & 262094 & \textbf{74599}\\ \midrule
		\multirow{3}{*}{set 1}       & PSNR  & 16.140  & 17.996 & \textcolor{blue}{20.574}  &  {20.074}        &  \textbf{21.304}            \\ 
		& SSIM  &  0.556   & 0.661 & \textcolor{blue}{0.873}   & 0.772  & \textbf{0.912}             \\ 
		& RMSE  &  0.155   &  0.1259  & \textcolor{blue}{0.094}         & {0.099}  & \textbf{0.086}          \\ \midrule
		\multirow{3}{*}{set 2}       & PSNR  & 14.975    & 16.508   &  \textcolor{blue}{19.432}         & 19.215     &  \textbf{20.013}          \\  
		& SSIM  &  0.472   &  0.630  & \textcolor{blue}{0.841}         & 0.725     & \textbf{0.882}          \\  
		& RMSE  &  0.178   &  0.149  & \textcolor{blue}{0.106}         & 0.109     & \textbf{0.099}          \\ \bottomrule
\end{tabular}}
\label{tab:metric_emt}
\end{table*}

\begin{figure}[!t]
\centerline{\includegraphics[width=2in]{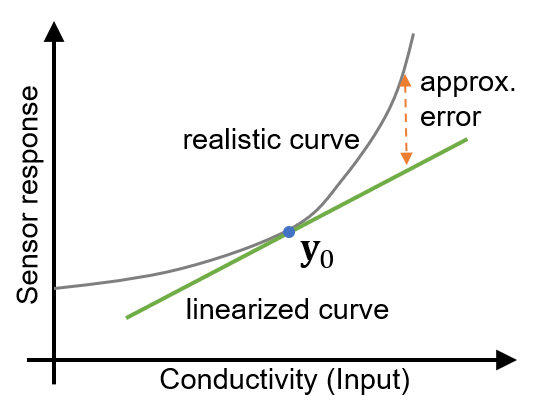}}
\caption{Sensor response of EMT. The nonlinear characteristic leads to absolute conductivity errors.}
\label{fig:emt_nonlinear}
\end{figure}

It is however worth mentioning that all methods failed to accurately reconstruct the absolute conductivity of different objects. One reason is that EMT image reconstruction is a nonlinear problem, where $\mathbf{A}_{\mathbf{y}}$ in (\ref{eq:basic}) is the sensitivity matrix of coils over spatial conductivity in the sensing region; $\mathbf{y}$ is determined by the true conductivity $\mathbf{x}$, spatial coordinate, excitation signals, etc. $\mathbf{A}_{\mathbf{y}}(\cdot)$ is customarily linearized around a constant $\mathbf{y}_0$, leading to an approximated linear problem (see Fig. \ref{fig:emt_nonlinear}). Nonlinear errors arising from linear approximation account for the imprecise absolute conductivity in the reconstructed images.  Though the forward operator $\mathbf{A}_{\mathbf{y}}(\cdot)$ can be updated during iteration to achieve nonlinear reconstruction, the computation cost and modeling errors using finite element method make it prohibitive especially in time critical applications.

\subsubsection{Effect of Number of Layers}
\label{sec:layers}
To evaluate the effect of the number of layers, we tuned the value from 5 to 9. Table \ref{tab:emt_layers} summarizes the quantitative evaluation results of FISTA-Net on the validation set. The performance is gradually improved with the increase of the number of layers. Then the performance {tends to reach a stable level} when the layer number is larger than 7. Although the number of learnable parameters is fixed in all configurations, the GPU computational cost increases considerably when the number of layers is larger than 9. Based on observation, the 7-layer configuration is a preferable setting compromising the image quality and computational cost.

\begin{table}[!tb]
\caption{{Quantitative results of different number of layers}}
\centering{
	\begin{tabular}{@{}llllll@{}}
		\toprule
		Layers & 5 & 6 & {7} & 8 & 9 \\ \midrule
		PSNR         & 19.647  & 19.825  & {20.013}  & 20.135  &  20.318  \\
		SSIM         & 0.870  & 0.876  & {0.882}  &  0.885  &  0.889  \\
		RMSE         & 0.104  & 0.102  & {0.099}  &  0.099  &  0.096  \\ \bottomrule
\end{tabular}}
\label{tab:emt_layers}
\end{table}

\subsubsection{Iteration results}
\label{subsub:iterative_emt}

\begin{figure}[tbp]
\centering
\subfigure[FISTA-Net iteration results (starting point from $\mathbf{X}_0$).]{\label{subfig:emt_iteration_fistanet}\includegraphics[width =3 in]{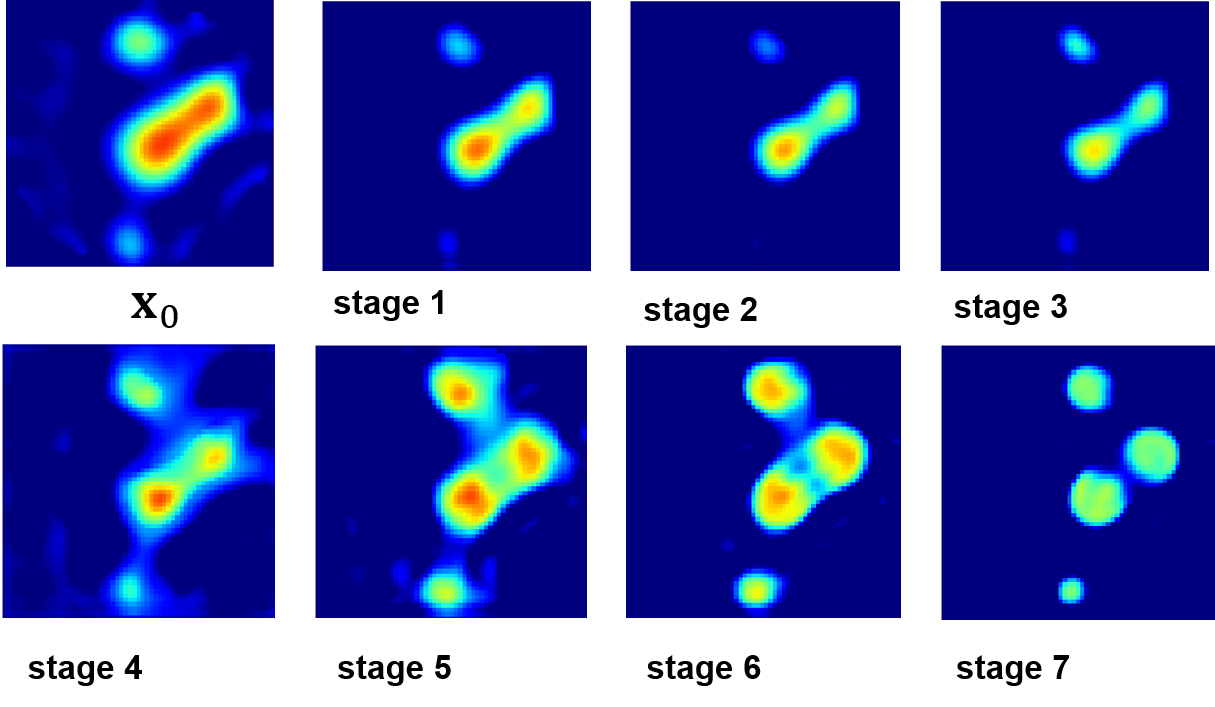}}
\hspace{0.1in}
\subfigure[FISTA-TV\cite{Beck2009FastGA} iteration results (starting point from zero).]{\includegraphics[width =3 in]{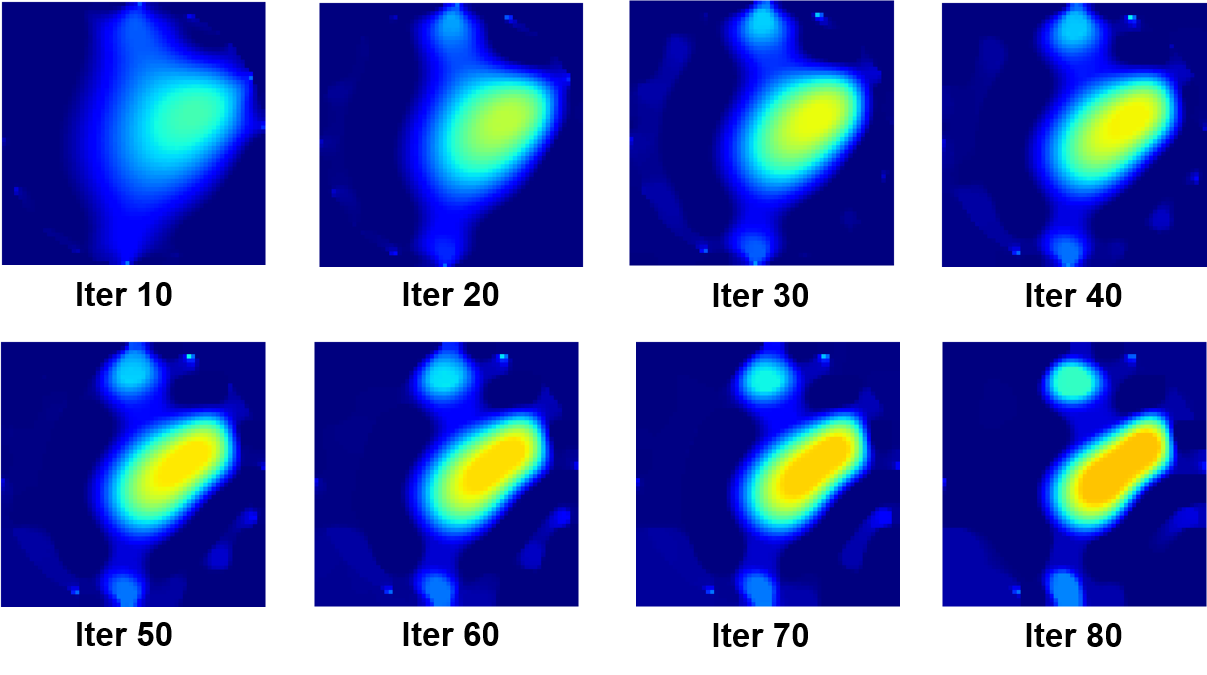}}
\caption{Reconstructed intermediate EMT images by FISTA-Net and FISTA-TV at different stages.}
\label{fig:emt_iteration}
\end{figure}

\begin{figure}[tbp]
\centering
\subfigure[FISTA-Net learned parameters]{\includegraphics[width =1.55 in]{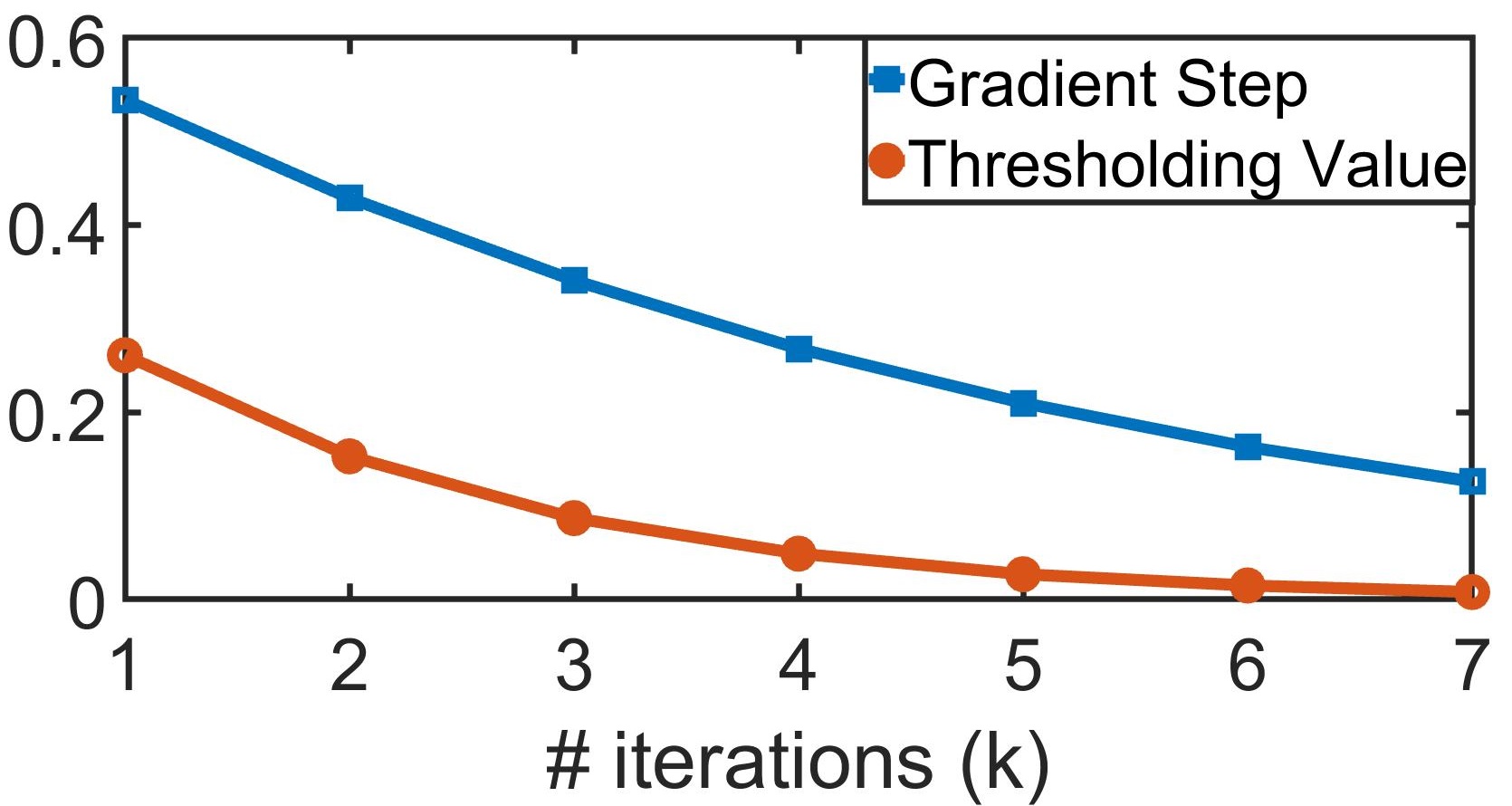}}
\hspace{0in}
\subfigure[ISTA-Net learned parameters]{\label{subfig:istanet_pars}\includegraphics[width =1.6 in]{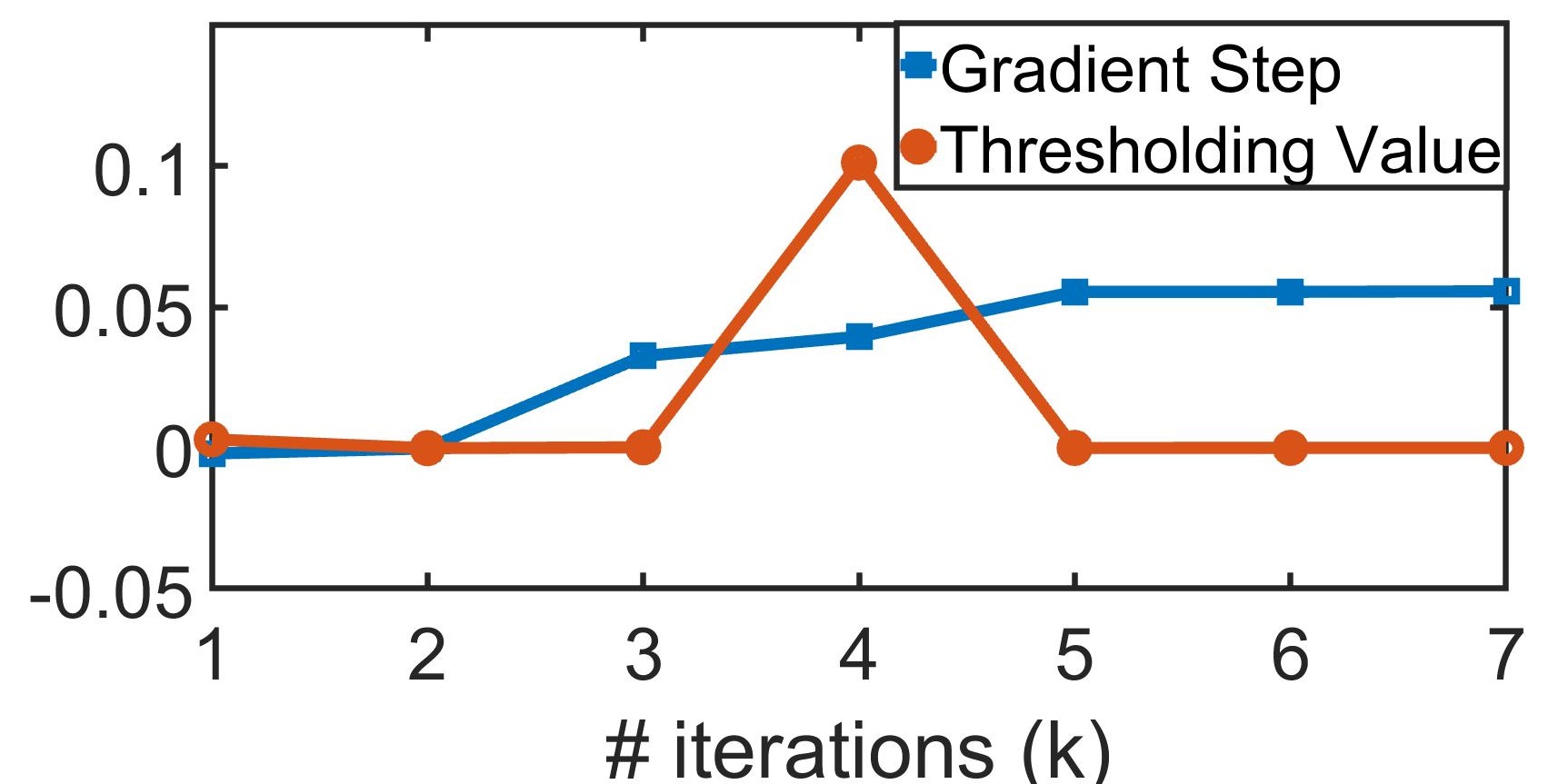}}
\caption{{Learned parameters of (a) FISTA-Net and (b) ISTA-Net with respect to different iteration step $k$.}}
\label{fig:iteration_pars}
\end{figure}

{Fig. \ref{fig:emt_iteration} and \ref{fig:iteration_pars} show the intermediate images and learned parameters of FISTANet and ISTA-Net during iteration. In Fig. \ref{subfig:emt_iteration_fistanet}, the intermediate images from layer 1-7 of FISTA-Net become gradually clearer, and the model-based parameters  $\theta^{(k)}, \mu^{(k)}$ of FISTA-Net decrease monotonically  with the number of iterations. Smaller $\theta^{(k)}$ and $\mu^{(k)}$  lead to a more accurate solution but slower convergence. This accords with the definition of $\theta^{(k)}$ and $\mu^{(k)}$ which also implies that the noise variance of the reconstructed image is gradually decreasing. The phenomenon is consistent with the parameter configuration in the conventional model-based methods, e.g. adaptive LASSO which uses a large $\theta^{(k)}$ for the small $k$, and gradually reduces it as $k$ increases to improve the accuracy\cite{Hale2008FixedPointCF}.
Differently, in Fig. \ref{subfig:istanet_pars}, the learned parameters of ISTA-Net have poor interpretability without patterns. At some points, the parameters are negative, which contradicts their definition. On the contrary, the learned parameters of FISTA-Net are more meaningful.}

\subsubsection{Generalization Ability}
\begin{figure}[!t]
\centerline{\includegraphics[width=3.2in]{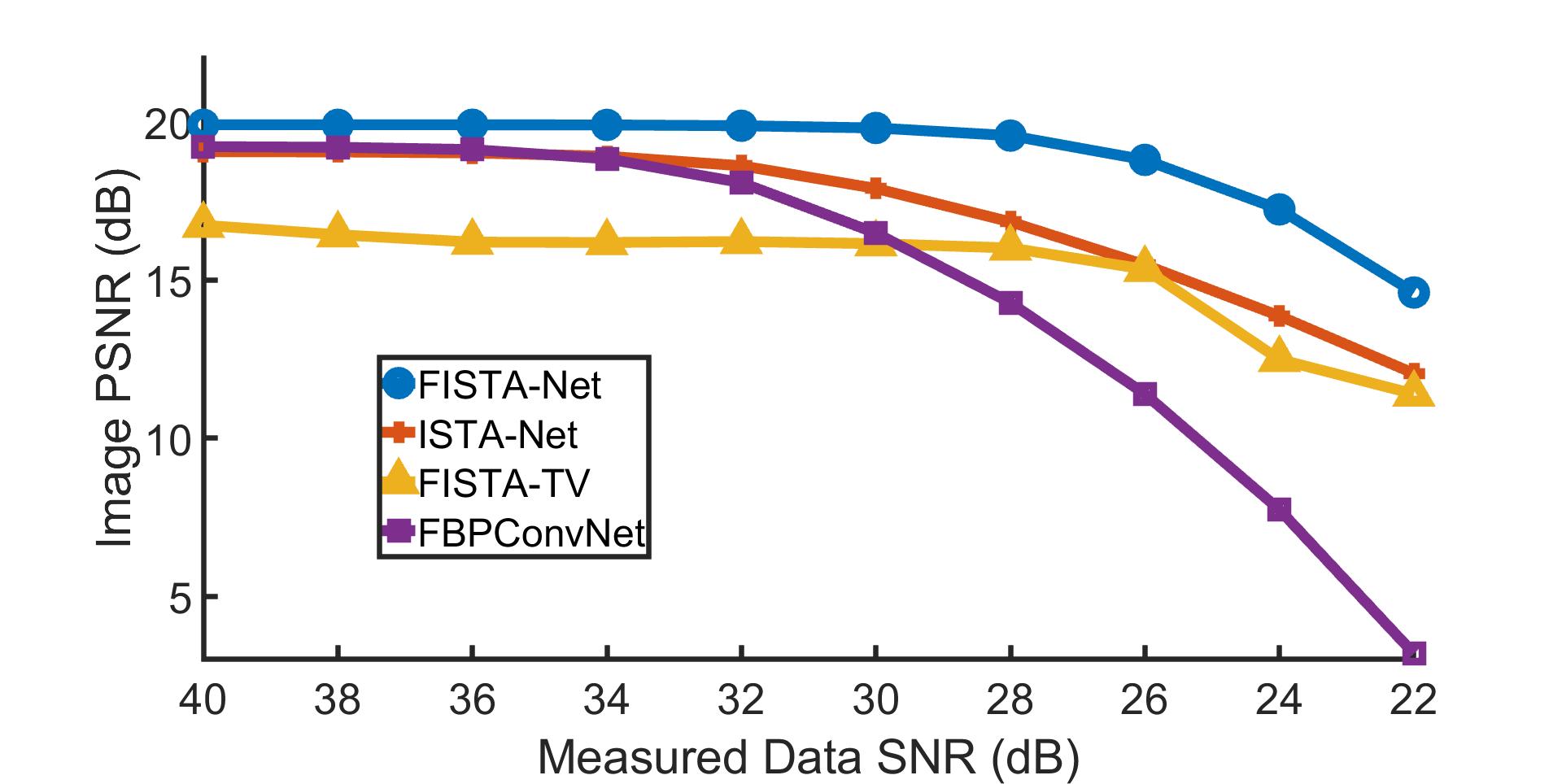}}
\caption{{Robustness study for additive noise in the test data. Normally distributed random noise is added to the measured data and the reconstruction quality is evaluated for all the algorithms under consideration.}}
\label{fig:robustness}
\end{figure}

{We conducted robustness studies for different noise levels (22dB to 40dB) to further evaluate the generalization ability of FISTA-Net. As the experimental data in realistic scenarios contain natural noise components, we added additional normally distributed noise to the measured data in the test set and recorded the PSNR of the reconstructed images (see Fig. \ref{fig:robustness}). The results show that: (1) FISTA-Net outperforms the other methods in the whole SNR range, which is more significant in low SNR settings; (2) model-based deep learning approaches, i.e. ISTA-Net and FISTA-Net, are more robust to additive noise, whereas the image post-processing approach, FBPConvNet, degrades much faster. As expected, FISTA-TV demonstrates similar robustness as ISTA-Net and FISTA-Net, because they inherit from the same physical prior of the forward operator.} 

\subsection{Linear Case: Sparse-view CT}

\subsubsection{Clinical CT dataset}
CT image reconstruction is a linear format of (\ref{eq:basic}), where $\mathbf{A}$ is the discretized Radon transform of the object under investigation. 
{A prevailing research thrust in recent years is to reduce the radiation dose in order to finally reduce the potential risk of radiation exposure. Among various approaches for low-dose CT, sparse-view CT is a recent proposal to lower the radiation dose by reducing the number of projection views. It can also reduce the scan time for continuous acquisition.} 
Clinical CT data and images established by Mayo Clinics, i.e. '2016 NIH-AAPM-Mayo Clinic Low-Dose CT Great Challenge'\cite{McCollough2016TUFG207A04OO}, are used to evaluate the reconstruction performance. The image dataset contains 2,378 full-dose CT images of 3mm thickness from ten patients. Sinograms for this dataset are 729 pixels by 720 views and are created by re-projecting using the Matlab function \textit{radon}. The reference images were reconstructed by \textit{iradon} operator in MATLAB using all 720 views\cite{Han2018FramingUV, Jin2017DeepCN}. The projection data is down-sampled to 60 and 120 views, respectively, to simulate a few view geometries. Among the ten patients' data, eight patients' data were used for training and one patient's data for validation, and the remaining one for testing. In detail, there are 1639 slices of 512 $\times$ 512 images in the training set, and 409 slices of 512 $\times$ 512 images for validation. The test data contain 330 slices of 512 $\times$ 512 images.

\subsubsection{Comparison Study}

\begin{table*}[!t]
\caption{Comparison of FISTA-Net with the state of the art imaging approaches on CT dataset (60 view).}
\begin{spacing}{1}
	\centering
	\begin{tabular}{cccccc}
		Ground Truth & FBP ($\mathbf{x_0}$) & {FISTA-TV\cite{Beck2009FastGA}} & FBPConvNet\cite{Jin2017DeepCN} & {ISTA-Net\cite{Zhang2018ISTANetIO}} & \textbf{FISTA-Net}  \\
		\includegraphics[width=2.5cm]{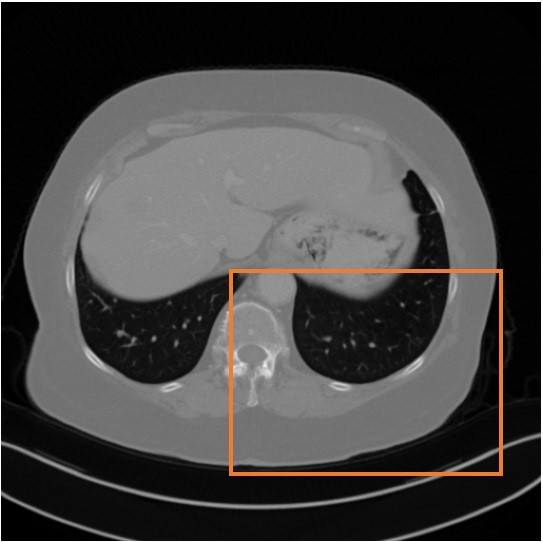} & 
		\includegraphics[width=2.5cm]{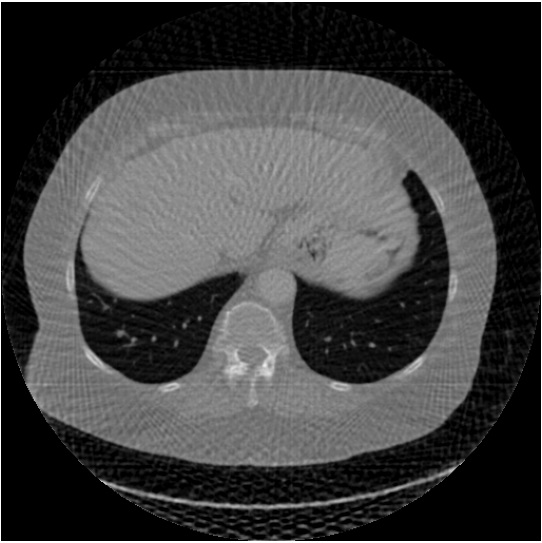} & 
		\includegraphics[width=2.5cm]{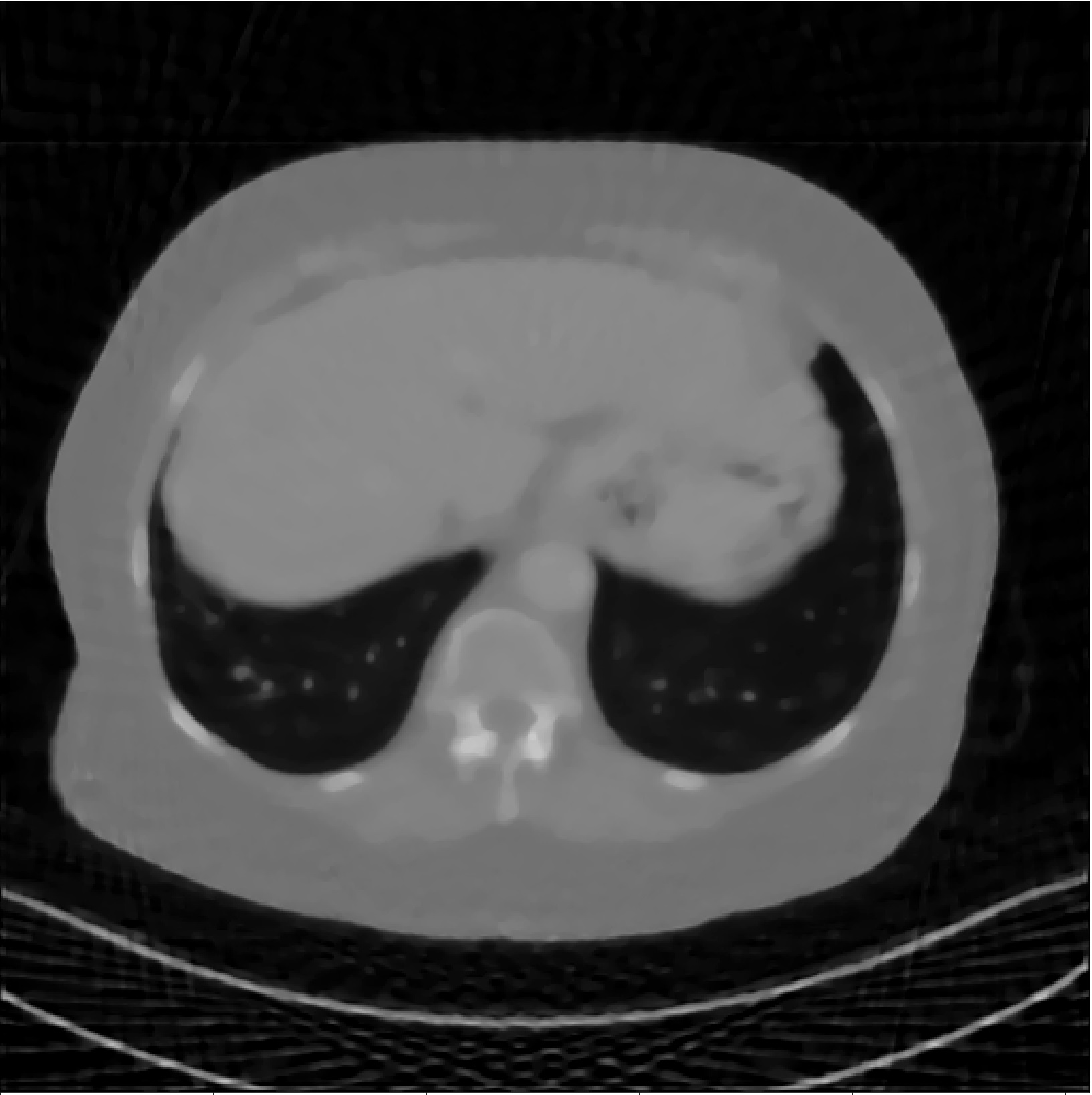} & 
		\includegraphics[width=2.5cm]{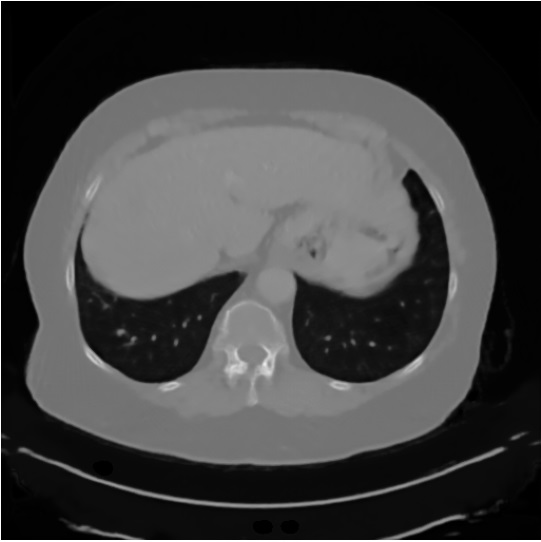} &
		\includegraphics[width=2.5cm]{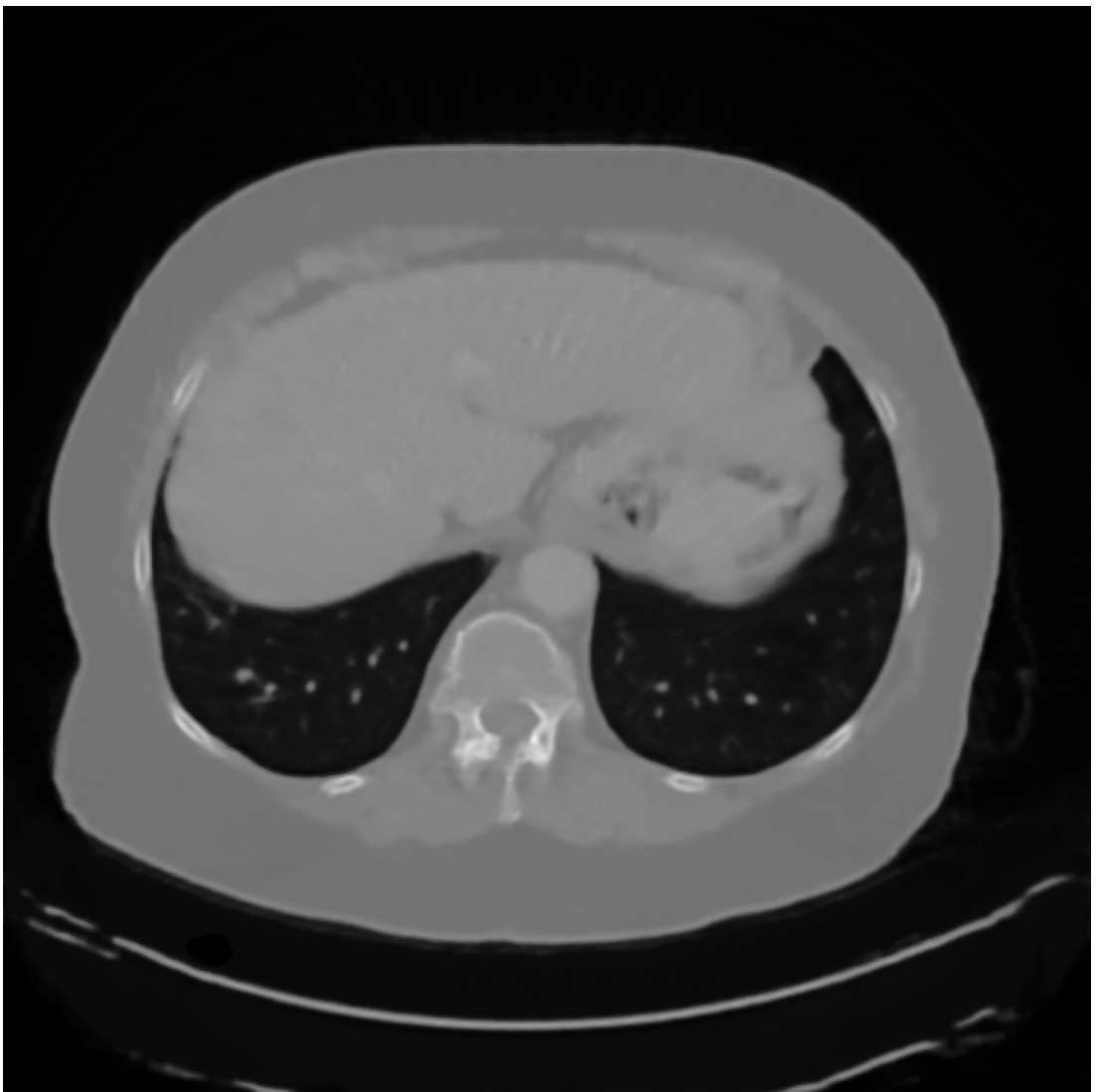} & 
		\includegraphics[width=2.5cm]{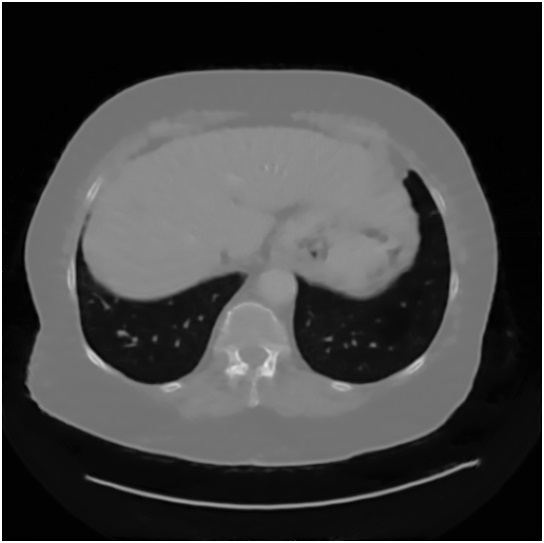} \\
		\includegraphics[width=2.5cm]{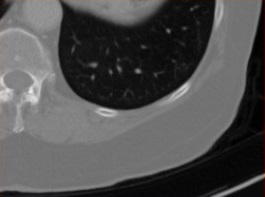} &
		\includegraphics[width=2.5cm]{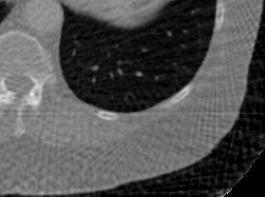} &
		\includegraphics[width=2.5cm]{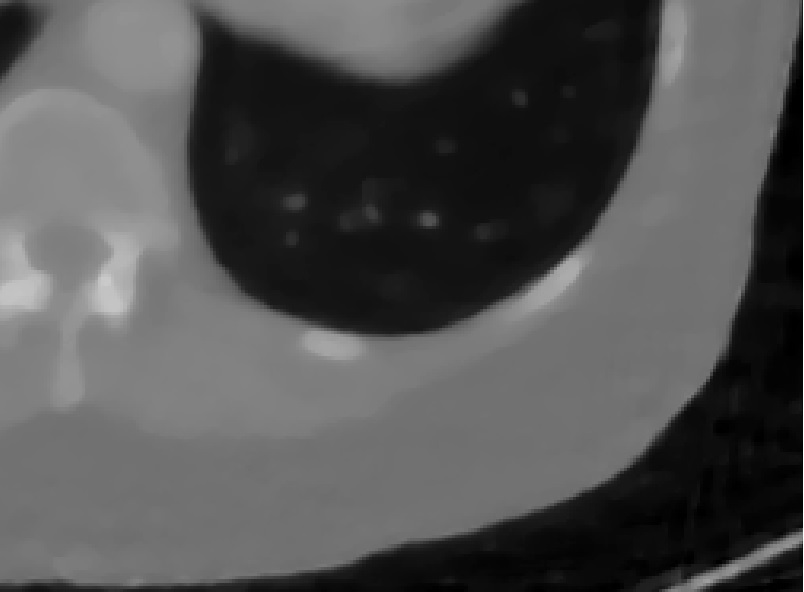} &
		\includegraphics[width=2.5cm]{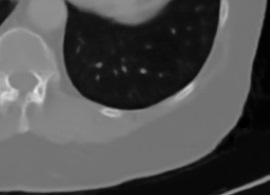} &
		\includegraphics[width=2.5cm]{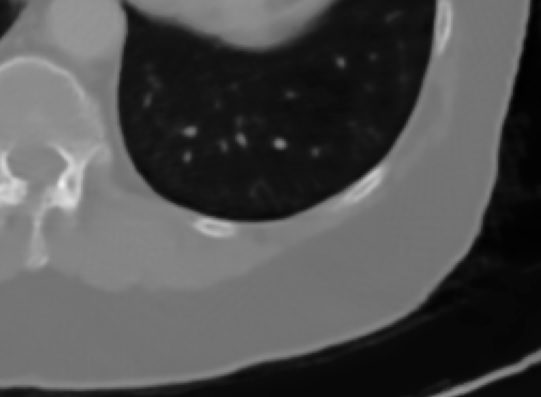} &
		\includegraphics[width=2.5cm]{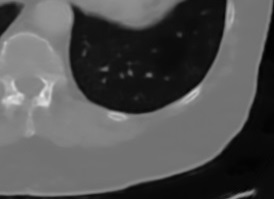} \\
		\includegraphics[width=2.5cm]{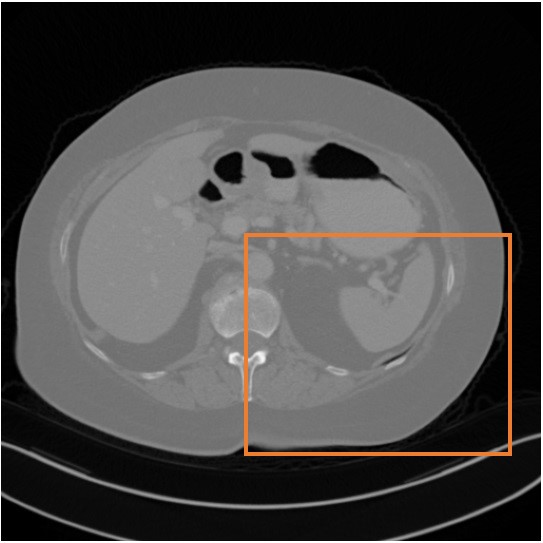} & 
		\includegraphics[width=2.5cm]{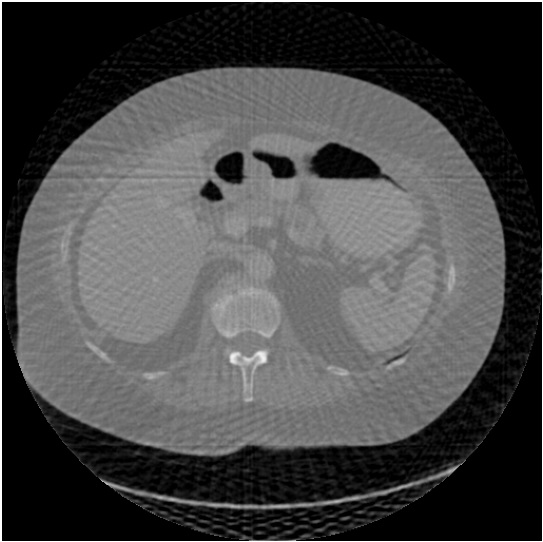} & 
		\includegraphics[width=2.5cm]{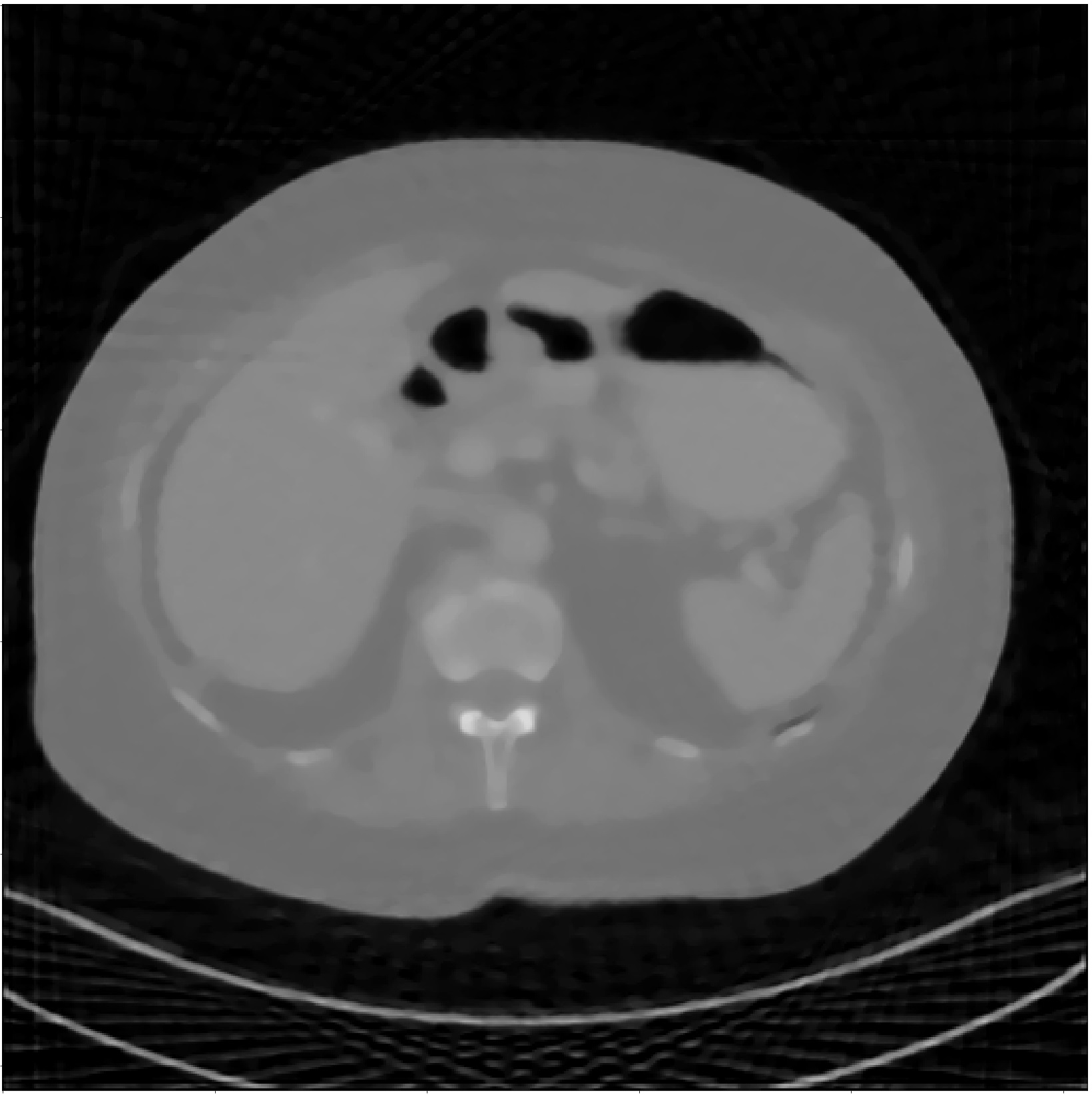} &
		\includegraphics[width=2.5cm]{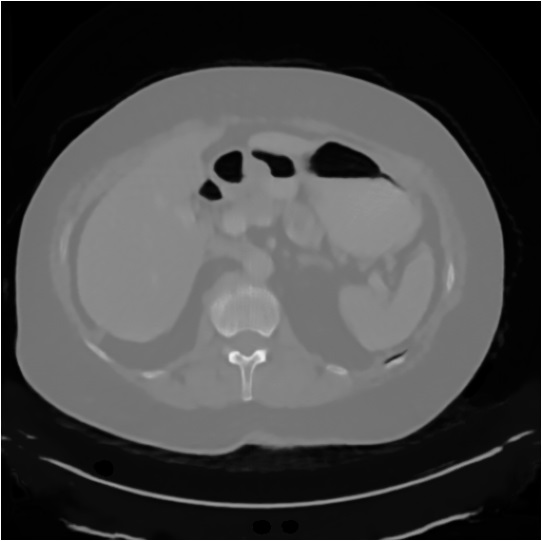} &
		\includegraphics[width=2.5cm]{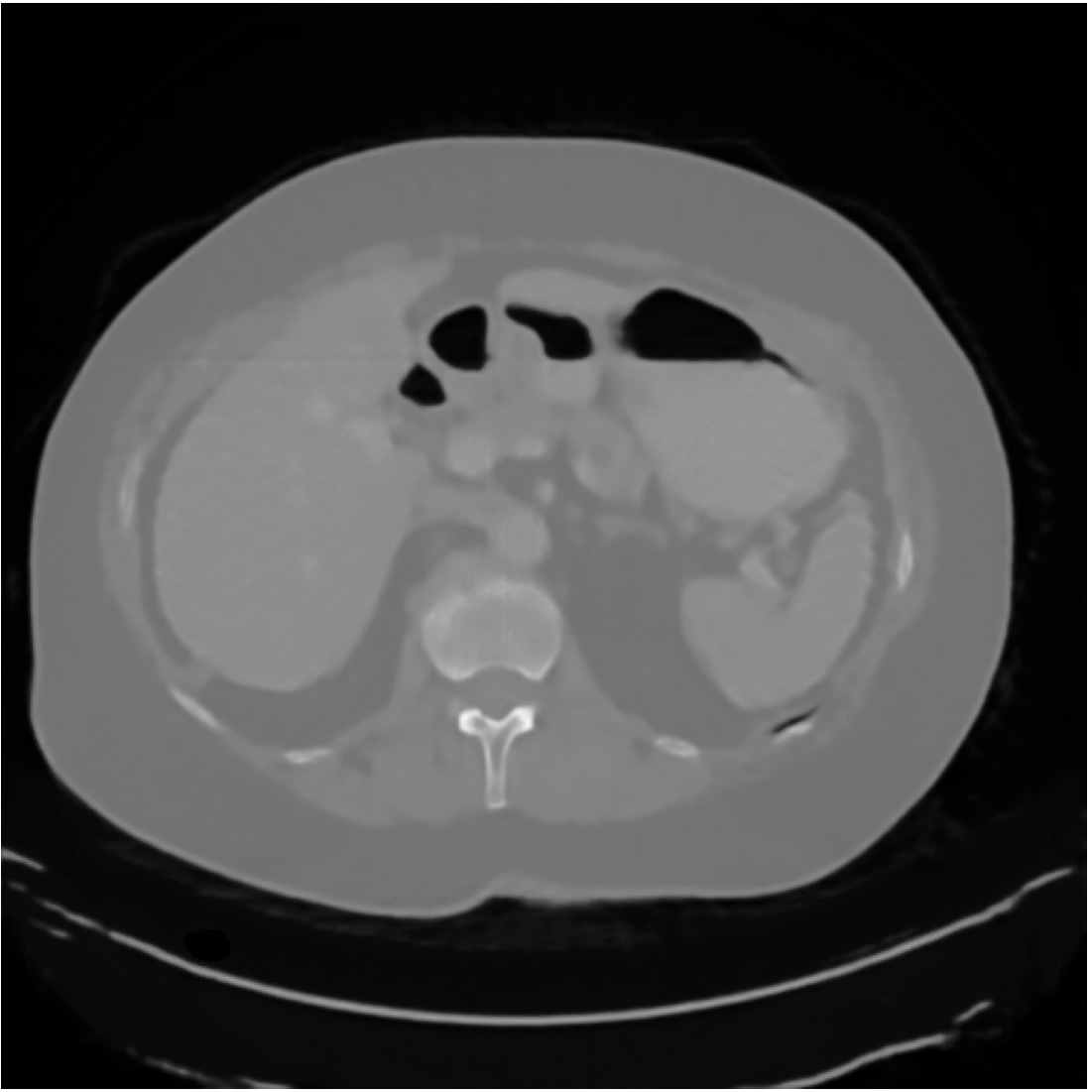} & 
		\includegraphics[width=2.5cm]{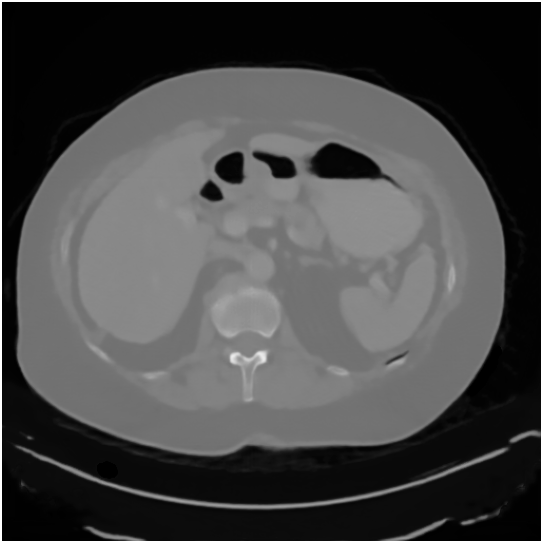} \\
		\includegraphics[height=2.2cm]{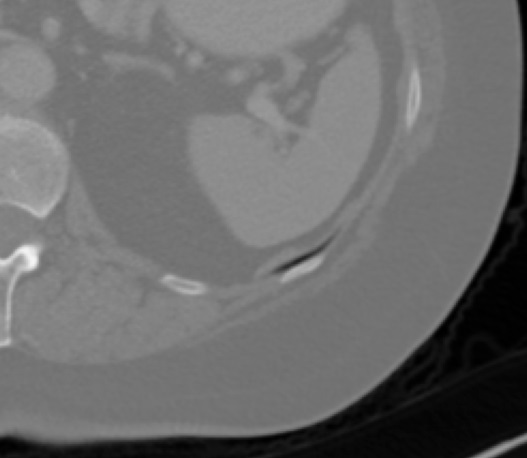} & 
		\includegraphics[height=2.2cm]{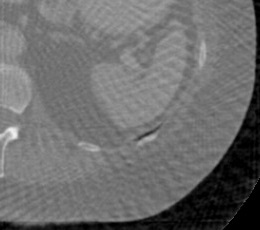} & 
		\includegraphics[height=2.2cm]{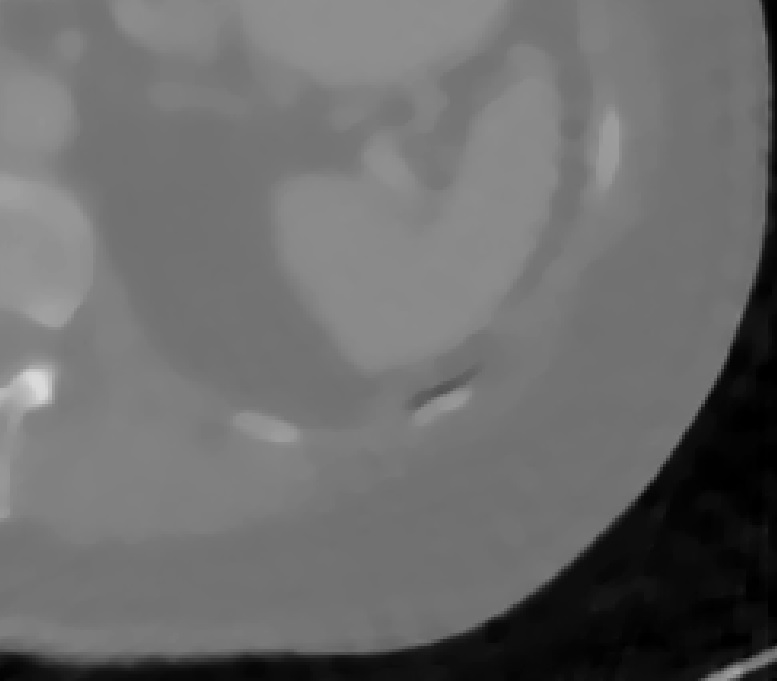} &
		\includegraphics[height=2.2cm]{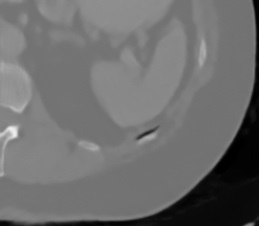} &
		\includegraphics[height=2.2cm]{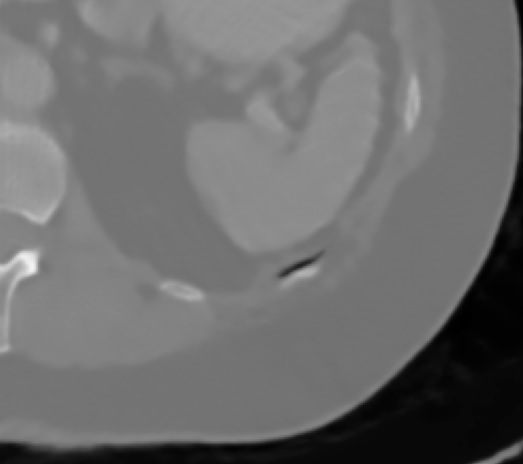} & 
		\includegraphics[height=2.2cm]{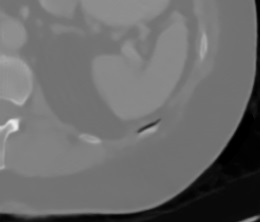} \\
	\end{tabular}
\end{spacing}
\label{tab:fig_ct}
\end{table*}

\begin{table*}[!t]
\caption{{Quantitative metrics of different methods on Mayo Clinic CT Dataset. The best results of comparative methods are highlighted in \textcolor{blue}{blue}}.}
\centering{
	\begin{tabular}{@{}cm{1.5cm}m{2.5cm}m{2.5cm}m{2.7cm}m{2.2cm}m{2cm}@{}}
		\toprule
		Testing set & Index	 & FBP ($\mathbf{x_0}$) & {FISTA-TV\cite{Beck2009FastGA}} & FBPConvNet\cite{Jin2017DeepCN} & {ISTA-Net\cite{Zhang2018ISTANetIO}} & \textbf{FISTA-Net}   \\ \midrule
		& \# of pars & 0 & 1 & 482449 & 262094 & \textbf{74599}\\ \midrule
		\multirow{3}{*}{60 view}  	& PSNR  	& 	27.533  &  33.251	&  	\textcolor{blue}{33.976}  	&  33.734   	&  \textbf{37.319}       	\\ 
		& SSIM  	&  	0.704	&  0.866	& 	\textcolor{blue}{0.942}   	&  0.920 		&  \textbf{0.951}        	\\ 
		& RMSE  	&  	0.042   &  0.021  	& 	\textcolor{blue}{0.020}     &   0.021		&  \textbf{0.013}        	\\ \midrule
		\multirow{3}{*}{120 view}   & PSNR  	&   29.152  &  35.217  	&  \textcolor{blue}{36.091}        &  35.452   		&   \textbf{40.189}        \\  
		& SSIM  	&   0.815  	&  0.949  	&   \textcolor{blue}{0.950}       &    0.948  		&   \textbf{0.968}        \\  
		& RMSE  	&   0.035  	&  0.017  	&   \textcolor{blue}{0.157}        &   0.016   		&   \textbf{0.009}       	\\ \bottomrule
\end{tabular}}
\label{tab:metric_ct}
\end{table*}

\begin{figure}[!t]
\centerline{\includegraphics[width=3.4in]{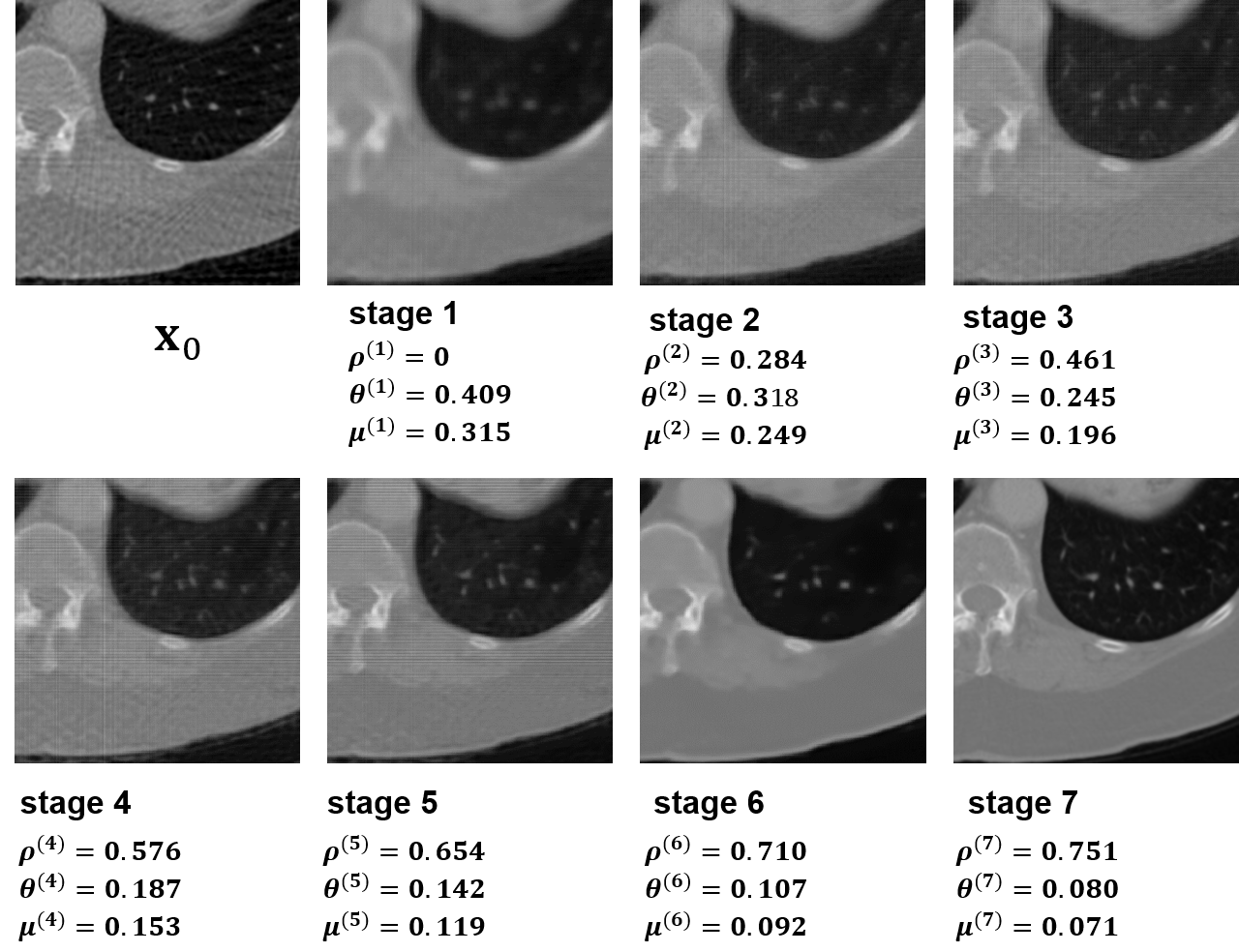}}
\caption{Reconstructed intermediate CT images (60-view ) by FISTA-Net at different stages and corresponding learned parameters.}
\label{fig:ct_iteration}
\end{figure}

Four state-of-the-art methods were compared against FISTA-Net, including FBP, FISTA-TV, FBP-ConvNet, and ISTA-Net. FBP is implemented with \textit{iradon} function in Matlab/Python. We set the maximum iteration steps of FISTA-TV as 100 and the regularization parameter to be 0.001. We set the cascaded stage of ISTA-Net as 6. Similarly, the number of layers of our FISTA-Net is configured as 7, following the same procedure described in Section \ref{sec:layers}.

Table \ref{tab:fig_ct} shows the reconstructed images on the Mayo Clinic CT dataset (60 view). In Table \ref{tab:fig_ct}, the sparse-view FBP contains obvious strip artifacts. FISTA-TV effectively removes the strip artifacts but some blocky artifacts are introduced, which is a typical feature of TV regularization. Additionally, FISTA-TV could over-smooth some fine structures (see the zoomed-in figures in row 2 and 4). FBPConvNet is advantageous over FISTA-TV in terms of noise reduction and strip artifacts suppression. However, comparing to the ground truth, some details are blurred or distorted to a certain extent. This phenomenon might be attributed to multiple downsampling and upsampling of UNet and the image quality could be sub-optimal, as commented by \cite{Chen2018LEARNLE}. In comparison, FISTA-Net involves measurements into the processing procedure to ensure the image quality is optimal. It maintained majority of the details and fine structures with superior noise reduction.

Table \ref{tab:metric_ct} lists the quantitative evaluation results of the 60 view and 120 view CT using different methods. The proposed FISTA-Net achieved the best results in terms of all the metrics, which is consistent with visual observations. In the cases of 60 and 120 views, FISTA-Net gained improvements of 3.340 dB and 4.098dB PSNR over FBPConvNet respectively, which achieved the best quantitative results among the comparative methods.

\subsubsection{Iteration results}
\label{subsub:iterative_ct}
Fig. \ref{fig:ct_iteration} shows the intermediate results of FISTA-Net in different iterations for a CT image with a sparse view of 60. The model-based parameter $\rho^{(k)}$ increases with respect to increasing stages while $\theta^{(k)}, \mu^{(k)}$ decreases. The noise removal and detail recovery effects are performed gradually across the iterations. As a result, the end-to-end trained FISTA-Net has similar parameter configurations with the conventional model-based reconstruction method. Therefore, the learned parameters facilitate image enhancement over pure network methods with meaningful model-based parameters.

\subsection{Convergence Analysis}
\begin{figure}[tbp]
\centering
\subfigure[Image reconstruction error on nonlinear EMT dataset.]{\includegraphics[width =2.8 in]{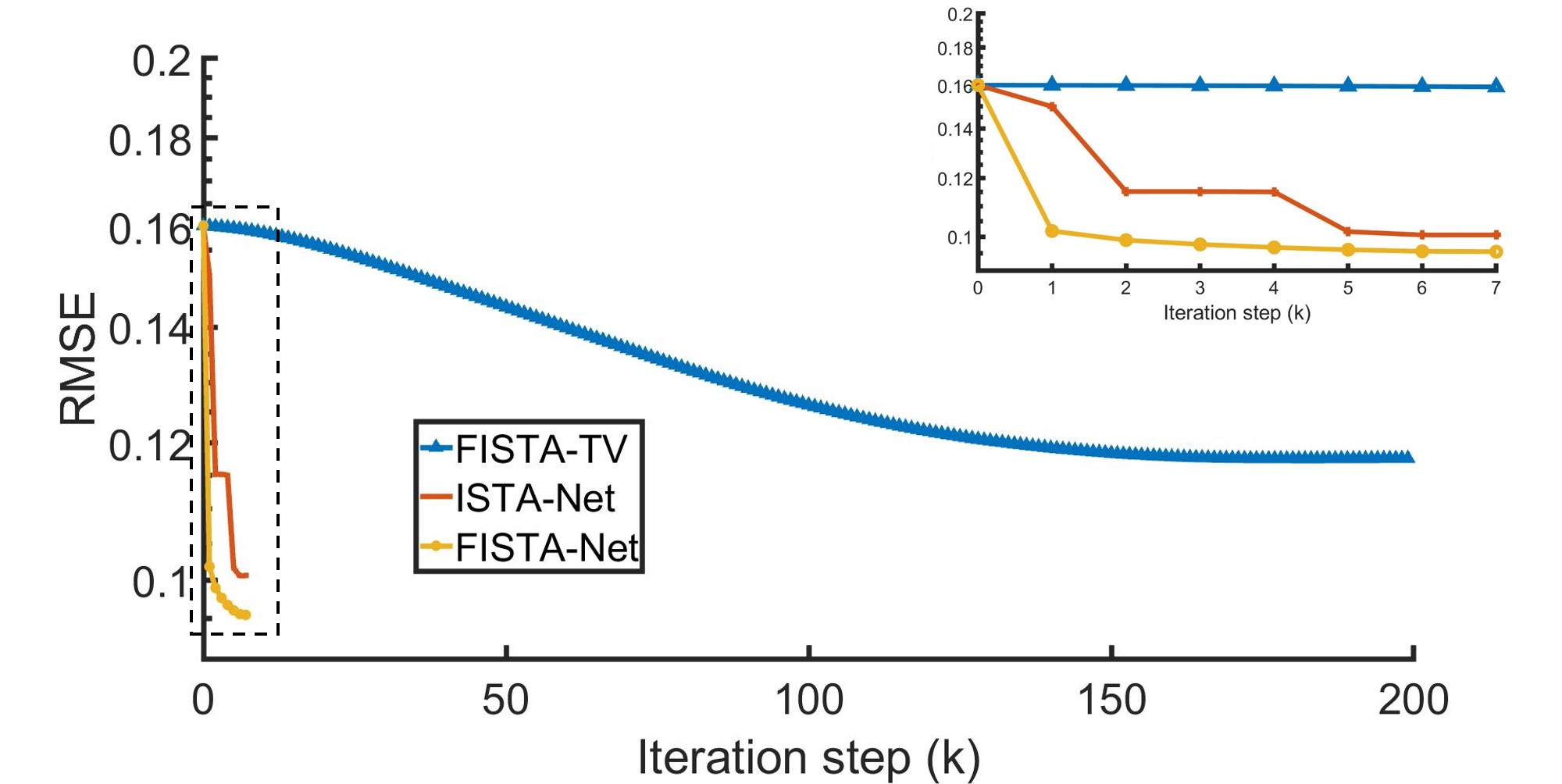}}
\hspace{0in}
\subfigure[Image reconstruction error on linear CT dataset.]{\includegraphics[width =2.8 in]{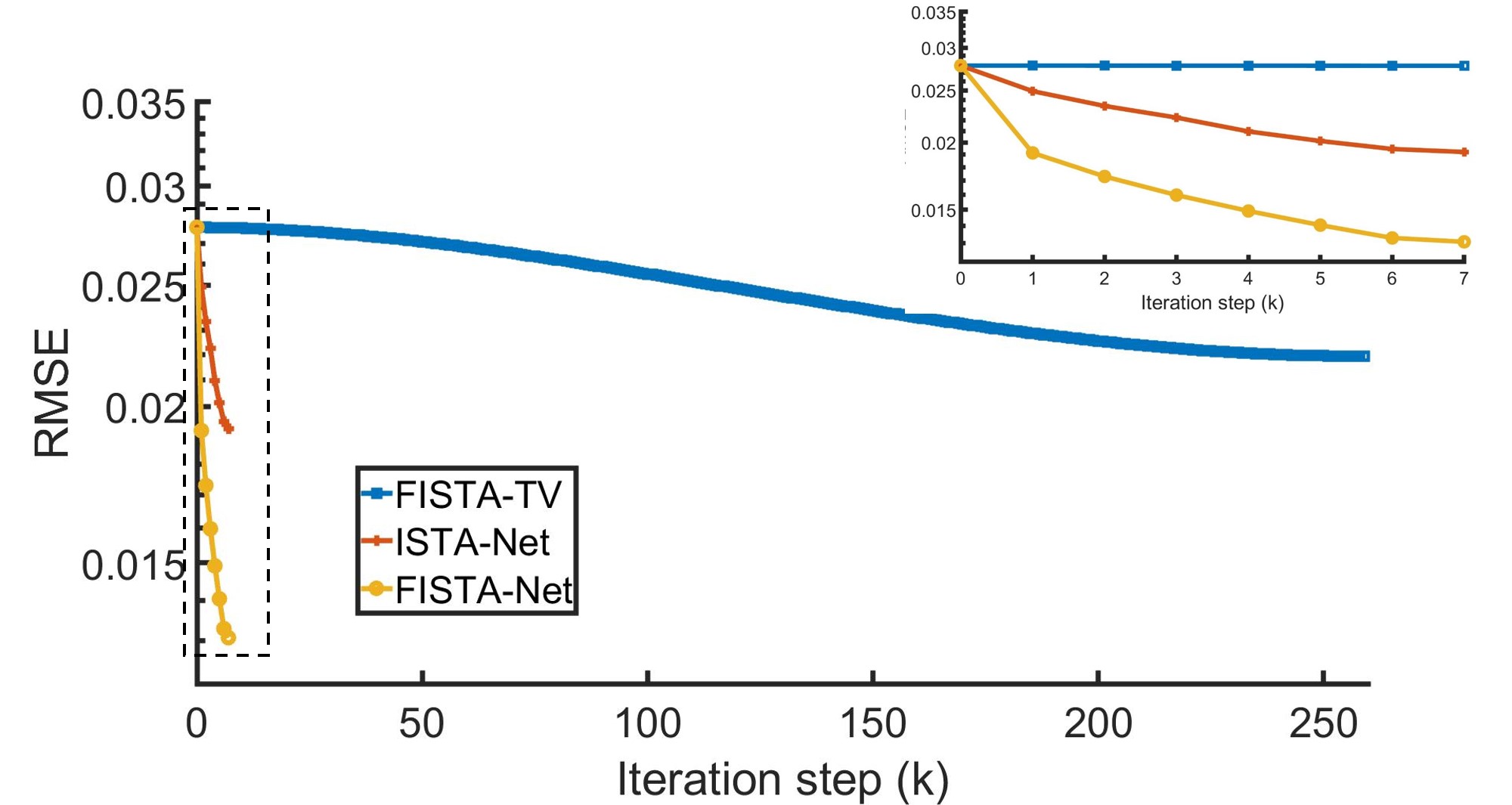}}
\caption{{Convergence performance analysis.}}
\label{fig:convergence}
\end{figure}

{The presented iterative results in Section \ref{subsub:iterative_emt} and  \ref{subsub:iterative_ct} provide intermediate images throughout iterations. In this section, we conduct more experimental analysis}
{concerning the convergence performance of FISTA-Net.}
{For comparison, we consider FISTA-TV, ISTA-Net and FISTA-Net using the same initial guess. The regularization parameter of FISTA-TV was hand-tuned to give the best reconstruction. Both ISTA-Net and FISTA-Net adopt 7 layers. Fig. \ref{fig:convergence} shows the evolution of the RMSE of FISTA-TV, ISTA-Net and FISTA-Net on both EMT and CT dataset. The conventional model-based FISTA-TV takes more than 200 iterations to converge to a stationary level whilst we only run the the model-based learning approaches, i.e. ISTA-Net and FISTA-Net, for 7 steps to obtain much better results. It can be observed that FISTA-Net converges faster than ISTA-Net and FISTA-TV. Also note that ISTA-Net converges in a stage-wise manner for the nonlinear EMT problem. This could be attributed to the freely learned parameters of ISTA-Net, which we discussed in} Section \ref{subsub:iterative_emt}.

{Rigorous theoretical analysis of convergence is not within the scope of this paper, but we refer the readers to the convergence analysis for Learned ISTA in\cite{Chen2018TheoreticalLC}, which can be potentially extended to FISTA-Net.}

\section{Conclusion}

We propose a model-based deep learning network named FISTA-Net to solve the inverse problems in imaging. FISTA-Net has good interpretability inherited from FISTA, and strong learning ability underpinned by CNN to perform noise removal and detail recovery. Through extensive experiments and analysis on two representative imaging modalities, i.e. EMT (non-linear case) and sparse-view CT (linear case), we demonstrated that: (1)  It shows good generalizability due to the decoupling of the data fidelity term and the proximal operator by unfolding FISTA into cascaded stages. (2) FISTA-Net can be readily extended to different inverse problems in imaging, by changing the physical model $\mathbf{A}$. 



\begin{thebibliography}{10}
\providecommand{\url}[1]{#1}
\csname url@samestyle\endcsname
\providecommand{\newblock}{\relax}
\providecommand{\bibinfo}[2]{#2}
\providecommand{\BIBentrySTDinterwordspacing}{\spaceskip=0pt\relax}
\providecommand{\BIBentryALTinterwordstretchfactor}{4}
\providecommand{\BIBentryALTinterwordspacing}{\spaceskip=\fontdimen2\font plus
	\BIBentryALTinterwordstretchfactor\fontdimen3\font minus
	\fontdimen4\font\relax}
\providecommand{\BIBforeignlanguage}[2]{{%
		\expandafter\ifx\csname l@#1\endcsname\relax
		\typeout{** WARNING: IEEEtran.bst: No hyphenation pattern has been}%
		\typeout{** loaded for the language `#1'. Using the pattern for}%
		\typeout{** the default language instead.}%
		\else
		\language=\csname l@#1\endcsname
		\fi
		#2}}
\providecommand{\BIBdecl}{\relax}
\BIBdecl

\bibitem{Chung2010AnEI}
J.~Chung and J.~G. Nagy, ``An efficient iterative approach for large-scale
separable nonlinear inverse problems,'' \emph{SIAM J. Scientific Computing},
vol.~31, pp. 4654--4674, 2010.

\bibitem{Lee2011CompressiveDO}
O.~K. Lee, J.~Kim, Y.~Bresler, and J.~C. Ye, ``Compressive diffuse optical
tomography: Noniterative exact reconstruction using joint sparsity,''
\emph{IEEE Transactions on Medical Imaging}, vol.~30, pp. 1129--1142, 2011.

\bibitem{figueiredo2003algorithm}
M.~A. Figueiredo and R.~D. Nowak, ``An em algorithm for wavelet-based image
restoration,'' \emph{IEEE Transactions on Image Processing}, vol.~12, no.~8,
pp. 906--916, 2003.

\bibitem{bioucas2007new}
J.~M. Bioucas-Dias and M.~A. Figueiredo, ``A new twist: Two-step iterative
shrinkage/thresholding algorithms for image restoration,'' \emph{IEEE
	Transactions on Image processing}, vol.~16, no.~12, pp. 2992--3004, 2007.

\bibitem{beck2009fast}
A.~Beck and M.~Teboulle, ``A fast iterative shrinkage-thresholding algorithm
for linear inverse problems,'' \emph{SIAM J. Imaging Sciences}, vol.~2, pp.
183--202, 2009.

\bibitem{Boyd2011DistributedOA}
S.~P. Boyd, N.~Parikh, E.~Chu, B.~Peleato, and J.~Eckstein, ``Distributed
optimization and statistical learning via the alternating direction method of
multipliers,'' \emph{Foundations and Trends in Machine Learning}, vol.~3, pp.
1--122, 2011.

\bibitem{Chambolle2010AFP}
A.~Chambolle and T.~Pock, ``A first-order primal-dual algorithm for convex
problems with applications to imaging,'' \emph{Journal of Mathematical
	Imaging and Vision}, vol.~40, pp. 120--145, 2010.

\bibitem{Kamilov2017APP}
U.~S. Kamilov, H.~Mansour, and B.~Wohlberg, ``A plug-and-play priors approach
for solving nonlinear imaging inverse problems,'' \emph{IEEE Signal
	Processing Letters}, vol.~24, pp. 1872--1876, 2017.

\bibitem{Wang2017NonconvexGO}
J.~Wang and L.~Zhao, ``Nonconvex generalizations of admm for nonlinear equality
constrained problems,'' \emph{ArXiv}, vol. abs/1705.03412, 2017.

\bibitem{Wei2020TuningfreePP}
K.~Wei, A.~I. Avil{\'e}s-Rivero, J.~Liang, Y.~Fu, C.-B. Sch{\"o}nlieb, and
H.~Huang, ``Tuning-free plug-and-play proximal algorithm for inverse imaging
problems,'' \emph{ArXiv}, vol. abs/2002.09611, 2020.

\bibitem{Kamilov2016ARB}
U.~Kamilov, D.~Liu, H.~Mansour, and P.~Boufounos, ``A recursive born approach
to nonlinear inverse scattering,'' \emph{IEEE Signal Processing Letters},
vol.~23, pp. 1052--1056, 2016.

\bibitem{Wang2018ImageRI}
G.~Wang, J.~C. Ye, K.~Mueller, and J.~A. Fessler, ``Image reconstruction is a
new frontier of machine learning,'' \emph{IEEE Transactions on Medical
	Imaging}, vol.~37, pp. 1289--1296, 2018.

\bibitem{adler2017solving}
J.~Adler and O.~{\"O}ktem, ``Solving ill-posed inverse problems using iterative
deep neural networks,'' \emph{Inverse Problems}, vol.~33, no.~12, p. 124007,
2017.

\bibitem{argyrou2012tomographic}
M.~Argyrou, D.~Maintas, C.~Tsoumpas, and E.~Stiliaris, ``Tomographic image
reconstruction based on artificial neural network (ann) techniques,'' in
\emph{2012 IEEE Nuclear Science Symposium and Medical Imaging Conference
	Record (NSS/MIC)}.\hskip 1em plus 0.5em minus 0.4em\relax IEEE, 2012, pp.
3324--3327.

\bibitem{tan2018image}
C.~Tan, S.~Lv, F.~Dong, and M.~Takei, ``Image reconstruction based on
convolutional neural network for electrical resistance tomography,''
\emph{IEEE Sensors Journal}, vol.~19, no.~1, pp. 196--204, 2018.

\bibitem{Zheng2018AnAI}
J.~Zheng and L.~Peng, ``An autoencoder-based image reconstruction for
electrical capacitance tomography,'' \emph{IEEE Sensors Journal}, vol.~18,
pp. 5464--5474, 2018.

\bibitem{Xiao2018DeepLI}
J.~Xiao, Z.~Liu, P.~Zhao, Y.~Li, and J.~Huo, ``Deep learning image
reconstruction simulation for electromagnetic tomography,'' \emph{IEEE
	Sensors Journal}, vol.~18, pp. 3290--3298, 2018.

\bibitem{McCann2017ConvolutionalNN}
M.~T. McCann, K.~H. Jin, and M.~A. Unser, ``Convolutional neural networks for
inverse problems in imaging: A review,'' \emph{IEEE Signal Processing
	Magazine}, vol.~34, pp. 85--95, 2017.

\bibitem{Yang2020ADMMCSNetAD}
Y.~Yang, J.~Sun, H.~Li, and Z.~Xu, ``Admm-csnet: A deep learning approach for
image compressive sensing,'' \emph{IEEE Transactions on Pattern Analysis and
	Machine Intelligence}, vol.~42, pp. 521--538, 2020.

\bibitem{Adler2018LearnedPR}
J.~Adler and O.~{\"O}ktem, ``Learned primal-dual reconstruction,'' \emph{IEEE
	Transactions on Medical Imaging}, vol.~37, pp. 1322--1332, 2018.

\bibitem{Jin2017DeepCN}
K.~H. Jin, M.~T. McCann, E.~Froustey, and M.~A. Unser, ``Deep convolutional
neural network for inverse problems in imaging,'' \emph{IEEE Transactions on
	Image Processing}, vol.~26, pp. 4509--4522, 2017.

\bibitem{han2016deep}
Y.~S. Han, J.~Yoo, and J.~C. Ye, ``Deep residual learning for compressed
sensing ct reconstruction via persistent homology analysis,'' \emph{arXiv
	preprint arXiv:1611.06391}, 2016.

\bibitem{Chen2017LowDoseCW}
H.~Chen, Y.~Zhang, M.~Kalra, F.~Lin, Y.~Chen, P.~Liao, J.~liu Zhou, and
G.~Wang, ``Low-dose ct with a residual encoder-decoder convolutional neural
network,'' \emph{IEEE Transactions on Medical Imaging}, vol.~36, pp.
2524--2535, 2017.

\bibitem{shan2019competitive}
H.~Shan, A.~Padole, F.~Homayounieh, U.~Kruger, R.~D. Khera, C.~Nitiwarangkul,
M.~K. Kalra, and G.~Wang, ``Competitive performance of a modularized deep
neural network compared to commercial algorithms for low-dose ct image
reconstruction,'' \emph{Nature Machine Intelligence}, vol.~1, no.~6, pp.
269--276, 2019.

\bibitem{Yang2018LowDoseCI}
Q.~Yang, P.~Yan, Y.~Zhang, H.~Yu, Y.~Shi, X.~Mou, M.~K. Kalra, Y.~Zhang,
L.~Sun, and G.~Wang, ``Low-dose ct image denoising using a generative
adversarial network with wasserstein distance and perceptual loss,''
\emph{IEEE Transactions on Medical Imaging}, vol.~37, pp. 1348--1357, 2018.

\bibitem{Aggarwal2019MoDLMD}
H.~K. Aggarwal, M.~Mani, and M.~Jacob, ``Modl: Model-based deep learning
architecture for inverse problems,'' \emph{IEEE Transactions on Medical
	Imaging}, vol.~38, pp. 394--405, 2019.

\bibitem{Antun2020OnIO}
V.~Antun, F.~Renna, C.~Poon, B.~Adcock, and A.~C. Hansen, ``On instabilities of
deep learning in image reconstruction - does ai come at a cost?''
\emph{Proceedings of the National Academy of Sciences of the United States of
	America}, 2020.

\bibitem{hammernik2018learning}
K.~Hammernik, T.~Klatzer, E.~Kobler, M.~P. Recht, D.~K. Sodickson, T.~Pock, and
F.~Knoll, ``Learning a variational network for reconstruction of accelerated
mri data,'' \emph{Magnetic resonance in medicine}, vol.~79, no.~6, pp.
3055--3071, 2018.

\bibitem{Maier2019LearningWK}
A.~Maier, C.~Syben, B.~Stimpel, T.~W{\"u}rfl, M.~Hoffmann, F.~Schebesch, W.~Fu,
L.~Mill, L.~Kling, and S.~Christiansen, ``Learning with known operators
reduces maximum training error bounds,'' \emph{Nature machine intelligence},
vol.~1, pp. 373 -- 380, 2019.

\bibitem{Gregor2010LearningFA}
K.~Gregor and Y.~LeCun, ``Learning fast approximations of sparse coding,'' in
\emph{ICML}, 2010.

\bibitem{Zhang2018ISTANetIO}
J.~Zhang and B.~Ghanem, ``Ista-net: Interpretable optimization-inspired deep
network for image compressive sensing,'' \emph{2018 IEEE/CVF Conference on
	Computer Vision and Pattern Recognition}, pp. 1828--1837, 2018.

\bibitem{Gong2017IterativePI}
K.~Gong, J.~Guan, K.~Kim, X.~Zhang, J.~Yang, Y.~Seo, G.~E. Fakhri, J.~Qi, and
Q.~Li, ``Iterative pet image reconstruction using convolutional neural
network representation,'' \emph{IEEE Transactions on Medical Imaging},
vol.~38, pp. 675--685, 2017.

\bibitem{Moreau2017UnderstandingTS}
T.~Moreau and J.~Bruna, ``Understanding trainable sparse coding with matrix
factorization,'' \emph{arXiv: Machine Learning}, 2017.

\bibitem{Daubechies2003AnIT}
I.~Daubechies, M.~Defrise, and C.~D. Mol, ``An iterative thresholding algorithm
for linear inverse problems with a sparsity constraint,''
\emph{Communications on Pure and Applied Mathematics}, vol.~57, pp.
1413--1457, 2003.

\bibitem{Elad2006WhySS}
M.~Elad, ``Why simple shrinkage is still relevant for redundant
representations?'' \emph{IEEE Transactions on Information Theory}, vol.~52,
pp. 5559--5569, 2006.

\bibitem{Combettes2005SignalRB}
P.~L. Combettes and V.~R. Wajs, ``Signal recovery by proximal forward-backward
splitting,'' \emph{Multiscale Model. Simul.}, vol.~4, pp. 1168--1200, 2005.

\bibitem{Afonso2010FastIR}
M.~V. Afonso, J.~M. Bioucas-Dias, and M.~A.~T. Figueiredo, ``Fast image
recovery using variable splitting and constrained optimization,'' \emph{IEEE
	Transactions on Image Processing}, vol.~19, pp. 2345--2356, 2010.

\bibitem{GuerquinKern2011AFW}
M.~Guerquin-Kern, M.~Haeberlin, K.~P. Pruessmann, and M.~Unser, ``A fast
wavelet-based reconstruction method for magnetic resonance imaging,''
\emph{IEEE Transactions on Medical Imaging}, vol.~30, pp. 1649--1660, 2011.

\bibitem{vonesch2008fast}
C.~Vonesch and M.~Unser, ``A fast thresholded landweber algorithm for
wavelet-regularized multidimensional deconvolution,'' \emph{IEEE Transactions
	on Image Processing}, vol.~17, no.~4, pp. 539--549, 2008.

\bibitem{zhang2015exponential}
Y.~Zhang, Z.~Dong, P.~Phillips, S.~Wang, G.~Ji, and J.~Yang, ``Exponential
wavelet iterative shrinkage thresholding algorithm for compressed sensing
magnetic resonance imaging,'' \emph{Information Sciences}, vol. 322, pp.
115--132, 2015.

\bibitem{Wu2020SparseCW}
K.~Wu, Y.~Guo, Z.~Li, and C.~Zhang, ``Sparse coding with gated learned ista,''
in \emph{ICLR}, 2020.

\bibitem{Chen2018TheoreticalLC}
X.~Chen, J.~Liu, Z.~Wang, and W.~Yin, ``Theoretical linear convergence of
unfolded ista and its practical weights and thresholds,'' \emph{ArXiv}, vol.
abs/1808.10038, 2018.

\bibitem{Liu2019ALISTAAW}
J.~Liu, X.~Chen, Z.~Wang, and W.~Yin, ``Alista: Analytic weights are as good as
learned weights in lista,'' in \emph{ICLR}, 2019.

\bibitem{Boyd2006ConvexO}
S.~P. Boyd and L.~Vandenberghe, ``Convex optimization,'' \emph{IEEE
	Transactions on Automatic Control}, vol.~51, pp. 1859--1859, 2006.

\bibitem{wang2018esrgan}
X.~Wang, K.~Yu, S.~Wu, J.~Gu, Y.~Liu, C.~Dong, Y.~Qiao, and C.~Change~Loy,
``Esrgan: Enhanced super-resolution generative adversarial networks,'' in
\emph{Proceedings of the European Conference on Computer Vision (ECCV)},
2018, pp. 0--0.

\bibitem{zhang2018residual}
Y.~Zhang, Y.~Tian, Y.~Kong, B.~Zhong, and Y.~Fu, ``Residual dense network for
image super-resolution,'' in \emph{Proceedings of the IEEE conference on
	computer vision and pattern recognition}, 2018, pp. 2472--2481.

\bibitem{goodfellow2016deep}
I.~Goodfellow, Y.~Bengio, and A.~Courville, \emph{Deep learning}.\hskip 1em
plus 0.5em minus 0.4em\relax MIT press, 2016.

\bibitem{Zhang2017ImageRF}
L.~Zhang and W.~Zuo, ``Image restoration: From sparse and low-rank priors to
deep priors [lecture notes],'' \emph{IEEE Signal Processing Magazine},
vol.~34, pp. 172--179, 2017.

\bibitem{Zhang2017LearningDC}
K.~Zhang, W.~Zuo, S.~Gu, and L.~Zhang, ``Learning deep cnn denoiser prior for
image restoration,'' \emph{2017 IEEE Conference on Computer Vision and
	Pattern Recognition (CVPR)}, pp. 2808--2817, 2017.

\bibitem{Nesterov2005SmoothMO}
Y.~Nesterov, ``Smooth minimization of non-smooth functions,''
\emph{Mathematical Programming}, vol. 103, pp. 127--152, 2005.

\bibitem{Yang2017AME}
Y.~Yang, J.~Jia, S.~Smith, N.~Jamil, W.~Gamal, and P.~O. Bagnaninchi, ``A
miniature electrical impedance tomography sensor and 3-d image reconstruction
for cell imaging,'' \emph{IEEE Sensors Journal}, vol.~17, pp. 514--523, 2017.

\bibitem{Xiang2019DesignOA}
J.~Xiang, Y.~Dong, M.~Zhang, and Y.~Li, ``Design of a magnetic induction
tomography system by gradiometer coils for conductive fluid imaging,''
\emph{IEEE Access}, vol.~7, pp. 56\,733--56\,744, 2019.

\bibitem{Ronchetti2020TorchRadonFD}
M.~Ronchetti, ``Torchradon: Fast differentiable routines for computed
tomography,'' \emph{ArXiv}, vol. abs/2009.14788, 2020.

\bibitem{Glorot2010UnderstandingTD}
X.~Glorot and Y.~Bengio, ``Understanding the difficulty of training deep
feedforward neural networks,'' in \emph{AISTATS}, 2010.

\bibitem{Kingma2015AdamAM}
D.~P. Kingma and J.~Ba, ``Adam: A method for stochastic optimization,''
\emph{CoRR}, vol. abs/1412.6980, 2015.

\bibitem{Wang2004ImageQA}
Z.~Wang, A.~C. Bovik, H.~R. Sheikh, and E.~P. Simoncelli, ``Image quality
assessment: from error visibility to structural similarity,'' \emph{IEEE
	Transactions on Image Processing}, vol.~13, pp. 600--612, 2004.

\bibitem{Xiang2020MultifrequencyET}
J.~Xiang, Y.~Dong, and Y.~Yang, ``Multi-frequency electromagnetic tomography
for acute stroke detection using frequency-constrained sparse bayesian
learning.'' \emph{IEEE transactions on medical imaging}, vol.~PP, 2020.

\bibitem{Yang2019ScaffoldBased3C}
Y.~Yang, H.~Wu, J.~Jia, and P.~Bagnaninchi, ``Scaffold-based 3-d cell culture
imaging using a miniature electrical impedance tomography sensor,''
\emph{IEEE Sensors Journal}, vol.~19, pp. 9071--9080, 2019.

\bibitem{Hu2019ImageRF}
D.~lin Hu, K.~Lu, and Y.~Yang, ``Image reconstruction for electrical impedance
tomography based on spatial invariant feature maps and convolutional neural
network,'' \emph{2019 IEEE International Conference on Imaging Systems and
	Techniques (IST)}, pp. 1--6, 2019.

\bibitem{Beck2009FastGA}
A.~Beck and M.~Teboulle, ``Fast gradient-based algorithms for constrained total
variation image denoising and deblurring problems,'' \emph{IEEE Transactions
	on Image Processing}, vol.~18, pp. 2419--2434, 2009.

\bibitem{Hale2008FixedPointCF}
E.~T. Hale, W.~Yin, and Y.~Zhang, ``Fixed-point continuation for
l1-minimization: Methodology and convergence,'' \emph{SIAM J. Optim.},
vol.~19, pp. 1107--1130, 2008.

\bibitem{McCollough2016TUFG207A04OO}
C.~McCollough, ``Tu-fg-207a-04: Overview of the low dose ct grand challenge.''
\emph{Medical physics}, vol. 43 6, pp. 3759--3760, 2016.

\bibitem{Han2018FramingUV}
Y.~Han and J.~C. Ye, ``Framing u-net via deep convolutional framelets:
Application to sparse-view ct,'' \emph{IEEE Transactions on Medical Imaging},
vol.~37, pp. 1418--1429, 2018.

\bibitem{Chen2018LEARNLE}
H.~Chen, Y.~Zhang, Y.~Chen, J.~Zhang, W.~Zhang, H.~Sun, Y.~Lv, P.~Liao,
J.~Zhou, and G.~Wang, ``Learn: Learned experts? assessment-based
reconstruction network for sparse-data ct,'' \emph{IEEE Transactions on
	Medical Imaging}, vol.~37, pp. 1333--1347, 2018.

\end{thebibliography}

\end{document}